\documentclass[aps,prd,twocolumn,superscriptaddress,nofootinbib]{revtex4-1}

\usepackage{latexsym}
\usepackage{amsmath}
\usepackage{amssymb}
\usepackage{amsfonts}
\usepackage{bm}

\usepackage{color}

\usepackage{supertabular} 
\usepackage{placeins}
\usepackage{epsfig}
\usepackage{graphicx}

\begin{document}


\title{Fully-charm and -bottom pentaquarks in a Lattice-QCD inspired quark model}


\author{Gang Yang}
\email[]{yanggang@zjnu.edu.cn}
\affiliation{Department of Physics, Zhejiang Normal University, Jinhua 321004, China}

\author{Jialun Ping}
\email[]{jlping@njnu.edu.cn}
\affiliation{Department of Physics and Jiangsu Key Laboratory for Numerical Simulation of Large Scale Complex Systems, Nanjing Normal University, Nanjing 210023, P. R. China}

\author{Jorge Segovia}
\email[]{jsegovia@upo.es}
\affiliation{Departamento de Sistemas F\'isicos, Qu\'imicos y Naturales, \\ Universidad Pablo de Olavide, E-41013 Sevilla, Spain}


\date{\today}

\begin{abstract}
The fully-charm and -bottom pentaquarks, \emph{i.e.} $cccc\bar{c}$ and $bbbb\bar{b}$, with spin-parity quantum numbers $J^P=\frac{1}{2}^-$, $\frac{3}{2}^-$ and $\frac{5}{2}^-$, are investigated within a Lattice-QCD inspired quark model, which has already successfully described the recently announced fully-charm tetraquark candidate $X(6900)$, and has also predicted several other fully-heavy tetraquarks. A powerful computational technique, based on the Gaussian expansion method combined with a complex-scaling range approach, is employed to predict, and distinguish, bound, resonance and scattering states of the mentioned five-body system. Both baryon-meson and diquark-diquark-antiquark configurations, along with all of their possible color channels are comprehensively considered. Narrow resonances are obtained in each spin-parity channel for the fully-charm and -bottom systems. Moreover, most of them seems to be compact multiquarks whose wave-functions are dominated by either hidden-color baryon-meson or diquark-diquark-antiquark structure, or by the coupling between them.
\end{abstract}


\maketitle


\section{Introduction}
\label{sec:intro}

One century of fundamental research in atomic and nuclear physics has shown that all matter is corpuscular. The atoms that comprise us contain a dense nuclear core which is composed of protons and neutrons, referred to collectively as nucleons, which are members of a broader class of femtometre-scale particles, called hadrons. In our understanding of hadrons, we have discovered that they are complicated bound-states of quarks (and gluons) whose interactions are described by a quantum non-Abelian gauge field theory called Quantum Chromodynamics (QCD).

A very successful classification scheme for hadrons in terms of their valence quarks and antiquarks was independently proposed by Murray Gell-Mann~\cite{Gell-Mann:1964ewy} and George Zweig~\cite{Zweig:1964CERN} in 1964. This classification was called the quark model, it basically separates hadrons in two big families: mesons (quark-antiquark) and baryons (three-quarks). The quark model received experimental verification beginning in the late 1960s and, despite extensive experimental searches, no unambiguous candidates for exotic quark-gluon configurations were identified until 2003, when the Belle Collaboration discovered an unexpected enhancement at $3872\,\text{MeV}$ in the $\pi^{+}\pi^{-}J/\psi$ invariant mass spectrum while studying the reaction $B^{+}\to K^{+}\pi^{+}\pi^{-}J/\psi$~\cite{Belle:2003nnu}.

The so-called $X(3872)$ state challenged the quark model picture, leading to an explosion of related experimental activity since then. Consequently, more than twenty different charmonium- and bottomonium-like XYZ states have been reported by worldwide experimental collaborations. Very detailed reviews have been published on the current state of the subject; see for instance Refs.~\cite{Brambilla:2010cs, Brambilla:2014jmp, Olsen:2014qna, Esposito:2016noz, Chen:2016qju, Chen:2016spr, Karliner:2017qhf, Ali:2017jda, Guo:2017jvc, Liu:2019zoy, Yang:2020atz, Dong:2020hxe, Dong:2021bvy, Chen:2021erj, Meng:2022ozq}. In the last few years, more tetraquark candidates have been reported such as the charged charmonium-like tetraquark with strangeness $X_{0,1}(2900)$~\cite{LHCb:2020pxc, LHCb:2020bls}; $X(2600)$~\cite{BESIII:2022llk}; $X(1835)$, $X(2120)$ and $X(2370)$~\cite{BESIII:2021xoh}; $Z_{cs}(3985)$~\cite{BESIII:2020qkh}; $Z_{cs}(4000)$, $Z_{cs}(4220)$, $X(4630)$ and $X(4685)$~\cite{LHCb:2021uow}; $Y(4230)$ and $Y(4500)$~\cite{BESIII:2022joj}; the hidden charm structure $\psi_2(3823)$~\cite{LHCb:2022kpe, BESIII:2022yga}; and the doubly charmed tetraquark $T^+_{cc}$~\cite{LHCb:2021vvq, LHCb:2021auc}. In the baryon sector, experimental progress has been also developed; the first hidden-charm pentaquark $P^+_c(4380)$ was reported in 2015 by the LHCb collaboration~\cite{Aaij:2015tga}, and then more candidates were announced, \emph{e.g.}, the $P^+_c(4312)$, $P^+_c(4337)$, $P^+_c(4440)$ and $P^+_c(4457)$~\cite{lhcb:2019pc, LHCb:2021chn}, and also the hidden-charm pentaquark with strangeness $P^0_{cs}(4459)$~\cite{LHCb:2020jpq}.

In support of the experimental effort, theorists have been proposing different 
kinds of color-singlet clusters made by quarks, antiquarks and gluons which go beyond conventional mesons and baryons. In fact, at the birth of the quark model, Gell-Mann and Zweig indicated already that hadronic states with $qq\bar q\bar q$ and $qqqq\bar q$ content should exist in nature. More concretely, recent studies on tetraquarks that include mass spectrum, structure, decay and production properties can be found in, for instance, Refs.~\cite{Weinberg:2013cfa, Braaten:2014qka, Brodsky:2015wza, Chen:2016jxd, Eichten:2017ffp, Richard:2018yrm, Ortega:2018cnm, Bedolla:2019zwg, Ortega:2021xst, Esposito:2021ptx, Balassa:2021ssx}. Besides, many analysis on hidden-charm pentaquarks $P_c$ and $P_{cs}$ have been published in the last few years~\cite{Cheng:2021gca, Du:2021bgb, Yalikun:2021bfm, Wang:2021crr, Chen:2021obo, Du:2021fmt, Xie:2022hhv, Wang:2022oof, Deng:2022vkv, Park:2022nza, Chen:2022onm, Burns:2021jlu, Yang:2021pio, Shi:2021wyt, Li:2021ryu, Ling:2021lmq, Wu:2021gyn, Ruangyoo:2021aoi, Ling:2021sld, Lu:2021irg, Wu:2021caw, Du:2021fmf, Xiao:2021rgp, Zhu:2021lhd, Chen:2021tip, Yan:2021nio, Yang:2015bmv}. Meanwhile, several other types of pentaquark are theoretically studied, such as the hidden-strange and -bottom pentaquarks~\cite{Yang:2022uot, Huang:2021ave, Zhu:2020vto, Yang:2018oqd} as well as the single and doubly-charmed pentaquarks~\cite{Zhang:2020dwp, Xing:2022aij, Ozdem:2022vip, Xing:2021yid, Chen:2021kad, Chen:2021htr, gy:2020dcp}.

The fully-heavy tetraquark states $QQ\bar{Q}\bar{Q}$ ($Q=c,\,b$) have recently attracted much attention. In 2017, the CMS Collaboration reported a benchmark measurement of $\Upsilon(1S)$-pair production in $pp$ collisions at $\sqrt{s}$=8 TeV~\cite{CMS:2016liw} whose preliminary analysis seems to indicate an excess at $18.4\,\text{GeV}$ in the $\Upsilon(1S) \ell^+ \ell^-$ decay channel~\cite{Yi:2018fxo}. Besides, a significant peak at $18.2\,\text{GeV}$ was observed in Cu+Au collisions at RHIC~\cite{ANDY:2019bfn} but the LHCb and CMS collaborations~\cite{LHCb:2018uwm, CMS:2020qwa} were not able to confirm it from the $\Upsilon(1S)\mu^+ \mu^-$ invariant mass spectrum. Nevertheless, in the di-$J/\psi$ invariant mass spectrum, a narrow peak at 6.9 GeV, a broad one between 6.2 and 6.8 GeV, and a hint for a possible structure around 7.2 GeV were reported by the LHCb collaboration~\cite{LHCb:2020bwg}.

The identification of fully-heavy tetraquark states make one speculate that the pentaquark system consisting of all heavy quarks, \emph{i.e.} $QQQQ\bar Q$ with $Q$ either $c$- or $b$-quark, may also exist. The theoretical study of the masses and decay properties would help to search for the heavy pentaquark states in experiments. In fact, all investigations to date are as follows: the fully-charm pentaquark state should have a mass at around 7.9 GeV according to quark models~\cite{Yan:2021glh, An:2020jix, An:2022fvs} and QCD sum rules~\cite{Wang:2021xao, Zhang:2020vpz}; and it is claimed by the same works that the fully-bottom pentaquark should be located at approximately 23 GeV.

In view of the complexity of the problem at hand, the only way for progress is 
the use of a diverse array of theoretical approaches. In this work, we explore the possibility of having bound, resonance and scattering states of fully-charm and -bottom pentaquarks, \emph{viz.} $QQQQ\bar Q$ $(Q = c, b)$, with spin-parity $J^P=\frac{1}{2}^-$, $\frac{3}{2}^-$ and $\frac{5}{2}^-$ in the S-wave channel. We employ a non-relativistic quark model with a two-body interaction between heavy quarks based on the Lattice-QCD study of Ref.~\cite{Kawanai:2011jt} and successfully applied by us to the case of fully heavy tetraquarks in Ref.~\cite{Yang:2021hrb}. The five-body problem is solved by using the Gaussian expansion method~\cite{Hiyama:2003cu} combined with the complex scaling method~\cite{Aguilar:1971ve, Balslev:1971vb, Simon1972, HO19831} according to the so-called ABC theorem~\cite{Aguilar:1971ve, Balslev:1971vb}.

This work is organized in the following way. First, we briefly discuss the potential model, pentaquark wave functions, and computational method in Sec.~\ref{sec:model}. Then, Sec.~\ref{sec:results} is devoted to the analysis and discussion of our theoretical findings. A summary is followed in Sec.~\ref{sec:summary}.


\section{Theoretical framework}
\label{sec:model}

The nature of heavy quarks, either charm or bottom, can be well described within an non-relativistic quark model. Furthermore, inspired by Lattice-regularized QCD investigations such as Ref.~\cite{Kawanai:2011jt}, the interplay between a heavy quark, $Q$, and a heavy antiquark, $\bar{Q}$, can be well approximated by the spin-independent Cornell potential along with a spin-spin interaction~\cite{Eichten:1974af, Eichten:1979ms}. Accordingly, the Hamiltonian of five-heavy-flavored system can be generally expressed as
\begin{equation}
H = \sum_{i=1}^{5}\left( M_Q+\frac{\vec{p\,}^2_i}{2M_Q}\right) + \sum_{j>i=1}^{5} V(\vec{r}_{ij}) \,,
\label{eq:Hamiltonian}
\end{equation}
where $M_Q$ is the mass of heavy quark (antiquark), and the two-body interacting potential can be written as
\begin{align}
\label{CQMV}
V(\vec{r}_{ij}) = -\frac{3}{16}({\bf \lambda}^a_i\cdot{\bf \lambda}^a_j) & \Big[ -\frac{\alpha}{r_{ij}}+\sigma r_{ij} \nonumber \\
&
+ \beta e^{-\gamma r_{ij}}({\vec{s}}_{i}\cdot{\vec{s}}_{j}) \Big]  \,,
\end{align}
where, sort by appearance, it includes Coulomb, linear confining and spin-spin interactions. Moreover, color-dependence is encoded in the SU(3) Gell-Mann matrices, $\lambda^a_i$ $(a=1,2,...,8)$, and this factor is crucial in studying multiquark systems. Table~\ref{model} shows the model parameters: $\alpha$, $\beta$, $\gamma$, and $\sigma$; determined by fitting the conventional $Q\bar{Q}$ meson spectrum~\cite{Zhao:2020jqu, Yang:2021hrb}, and the theoretical and experimental data are summarized in Table~\ref{Mmeson}. One can see that  theoretical masses are consistent with their experimental counterparts. Although $\Omega_{ccc}$ and $\Omega_{bbb}$ baryons are still not seen by experiments, our calculated results are compatible with several other theoretical predictions~\cite{An:2022fvs, An:2020jix, Yang:2019lsg}. Therefore, this fact provides a solid ground to further study possible bound and resonance states in the fully-heavy pentaquark systems $QQQQ\bar{Q}$, and naturally continuing our work done for fully-heavy tetraquarks in Ref.~\cite{Yang:2021hrb}.

\begin{table}[!t]
\caption{\label{model} Model parameters.}
\begin{ruledtabular}
\begin{tabular}{llr}
Quark masses & $M_c$ $({\rm GeV})$        & 1.290 \\
             & $M_b$ $({\rm GeV})$        & 4.700 \\[2ex]
Coulomb      & $\alpha$                   & 0.4105 \\[2ex]
Confinement  & $\sigma$ $({\rm GeV}^2$)   & 0.2 \\[2ex]
Spin-Spin    & $\gamma$ $({\rm GeV})$     & 1.982 \\
             & $\beta_{cc}$ $({\rm GeV})$ & 2.06 \\
             & $\beta_{bb}$ $({\rm GeV})$ & 0.318 \\
\end{tabular}
\end{ruledtabular}
\end{table}

\begin{table}[!t]
\caption{\label{Mmeson} Theoretical and experimental masses, in MeV, for the $S$-wave conventional $Q\bar{Q}$ mesons and $QQQ$ baryons.}
\begin{ruledtabular}
\begin{tabular}{lcc}
State & $M_{\rm the.}$ & $M_{\rm exp.}$ \\
\hline
$\eta_c(1S)$       &  2968 & 2981 \\
$\eta_c(2S)$       &  3655 & 3639 \\
$\eta_c(3S)$       &  4152 & - \\
$J/\psi(1S)$       &  3102 & 3097 \\
$\psi(2S)$         &  3720 & 3686 \\
$\psi(3S)$         &  4200 & - \\ 
$\eta_b(1S)$       &  9401 & 9398 \\
$\eta_b(2S)$       &  9961 & 9999 \\
$\eta_b(3S)$       & 10316 & - \\
$\Upsilon(1S)$     &  9463 & 9460 \\
$\Upsilon(2S)$     &  9981 & 10023 \\
$\Upsilon(3S)$     & 10330 & 10355 \\
$\Omega_{ccc}(1S)$ &  4797 & - \\
$\Omega_{ccc}(2S)$ &  5305 & - \\
$\Omega_{ccc}(3S)$ &  5735 & - \\
$\Omega_{bbb}(1S)$ & 14358 & - \\
$\Omega_{bbb}(2S)$ & 14765 & - \\
$\Omega_{bbb}(3S)$ & 15083 & - \\
\end{tabular}
\end{ruledtabular}
\end{table}

\begin{figure}[!t]
\epsfxsize=3.4in \epsfbox{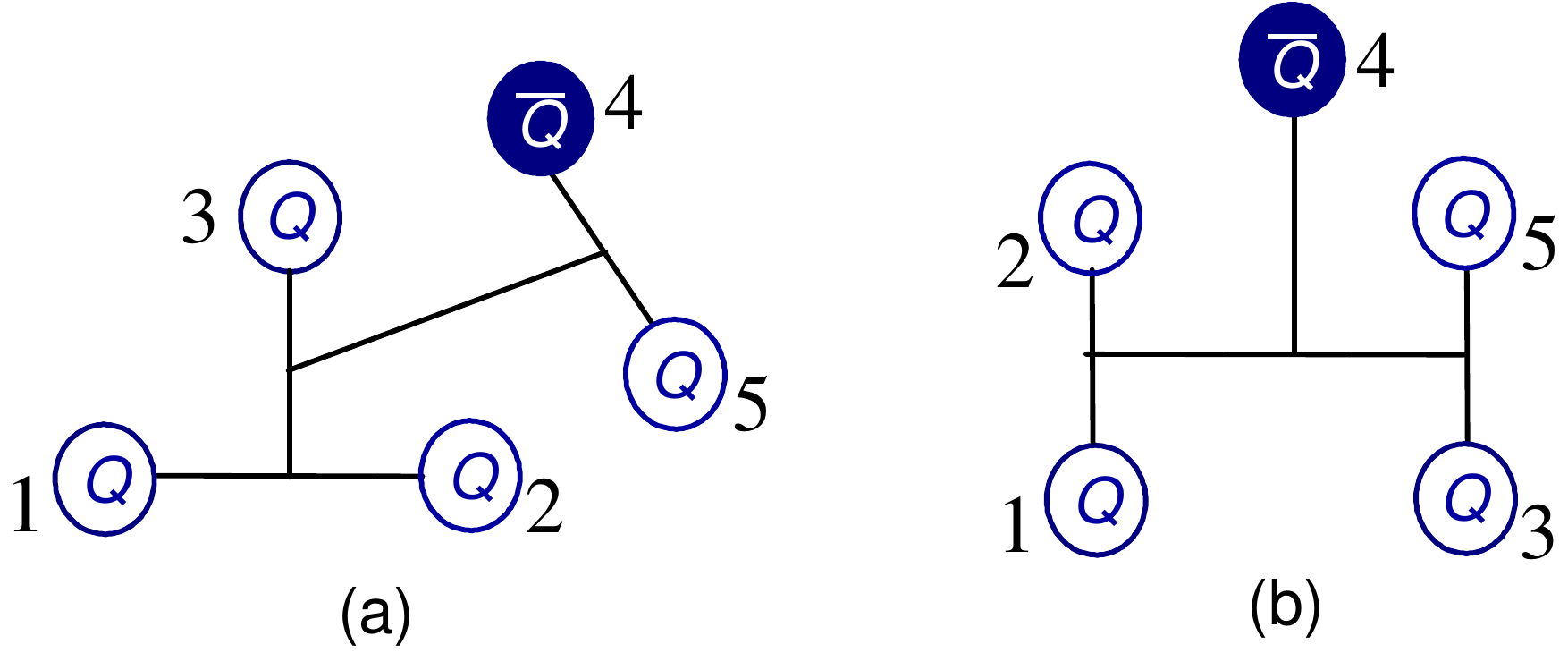}
\caption{Configurations in the fully-heavy pentaquarks. Panel $(a)$ is the baryon-meson structure, and panel $(b)$ is the diquark-diquark-antiquark one $(Q=c,b)$.} \label{FHPC}
\end{figure}

In the present investigation, both baryon-meson and diquark-diquark-antiquark configurations, shown in Fig.~\ref{FHPC}, are simultaneously considered. Moreover, we comprehensively include all possible color structures compatible with the mentioned 5-quark configurations and also allowed by the $S$-wave $J^P=1/2^-$, $3/2^-$ and $5/2^-$ quantum numbers studied. Additionally, different kinds of coupled-channel calculations have been performed: the case in which a particular quark arrangement configuration is considered, and the one where a complete coupled-channel computation is performed. Below, details on the fully-heavy pentaquark wave functions along with our computational approach are illustrated.

\subsection{Color, flavor an spin structure}

The pentaquark wave function is a product of four terms: color, flavor, spin and space. Concerning the color degree-of-freedom, there is a richer color-structure in multiquark systems than in conventional quark-antiquark and three-quark hadrons. Particularly, the colorless wave function of a $5$-body system can be obtained through either a color-singlet or a hidden-color channel or both. Since baryon-meson and diquark-diquark-antiquark configurations are simultaneously considered in this investigation, the color-singlet chanel can be reached considering only Fig.~\ref{FHPC}(a) whereas hidden-color channel can be obtained through either Fig.~\ref{FHPC}(a) or Fig.~\ref{FHPC}(b). Note herein that although it is enough to just consider the color singlet channel when all possible excited states of a system are included, a more efficient way of performing the computation is considering both the color singlet and hidden-color wave functions. The color-singlet wave function read as
\begin{align}
\label{Color1}
\chi^{c}_1 &= \frac{1}{\sqrt{18}}(rgb-rbg+gbr-grb+brg-bgr) \nonumber \\
&
\times (\bar r r+\bar gg+\bar bb) \,,
\end{align}
and the hidden-color one corresponding to Fig.~\ref{FHPC}(a) is
\begin{align}
\label{Color2}
\chi^{c}_k &= \frac{1}{\sqrt{8}}(\chi^{k}_{3,1}\chi_{2,8}-\chi^{k}_{3,2}\chi_{2,7}-\chi^{k}_{3,3}\chi_{2,6}+\chi^{k}_{3,4}\chi_{2,5} \nonumber \\
& +\chi^{k}_{3,5}\chi_{2,4}-\chi^{k}_{3,6}\chi_{2,3}-\chi^{k}_{3,7}\chi_{2,2}+\chi^{k}_{3,8}\chi_{2,1}) \,,
\end{align}
where $k=2\,(3)$ is an index which stands for the symmetric (anti-symmetric) configuration of two quarks in the $3$-quark sub-cluster. In Eq.~(\ref{Color2}), all color bases of the two sub-clusters are those used in studying the $P_c$ hidden-charm~\cite{Yang:2015bmv}, $P_b$ hidden-bottom~\cite{Yang:2018oqd}, and doubly charmed pentaquarks~\cite{gy:2020dcp}. Additionally, two colorless wave functions of diquark-diquark-antiquark configuration of Fig.~\ref{FHPC}(b) are obtained. Considering the following chains of color-coupling coefficients:\footnote{The group chain is obtained in sequence of quark number. Moreover, each quark and antiquark is represented with [1] and [11], respectively.}
\begin{itemize}
\item	\lbrack $C^{[2]}_{[1],[1]}C^{[11]}_{[1],[1]}C^{[211]}_{[2],[11]} C^{[222]}_{[11],[211]}$\rbrack$_4$, 

\item	\lbrack $C^{[11]}_{[1],[1]}C^{[11]}_{[1],[1]}C^{[211]}_{[11],[11]} C^{[222]}_{[11],[211]}$\rbrack$_5$,
\end{itemize}
the two colorless wave functions of diquark-diquark-antiquark configuration are (subscripts correspond to the ones above):
\begin{align}
\label{Color3}
\chi^{c}_4 &= \frac{1}{\sqrt{48}}\ \lbrace \bar{r}[(rb+br)(rg-gr)-(rg+gr)(rb-br)] + \nonumber \\
&
\bar{g}[(rg+gr)(gb-bg)+(gb+bg)(rg-gr)]+ \nonumber \\
&
\bar{b}[(rb+br)(gb-bg)-(gb+bg)(rb-br)]+ \nonumber \\
&
\sqrt{2}[\bar{r}rr(gb-bg)-\bar{g}gg(rb-br)+\bar{b}bb(rg-gr)] \rbrace\,,
\end{align}

\begin{align}
\label{Color4}
\chi^{c}_5 &= \frac{1}{\sqrt{24}}\ \lbrace \bar{r}[(rg-gr)(rb-br)-(rb-br)(rg-gr)] + \nonumber \\
&
\bar{g}[(rg-gr)(gb-bg)-(gb-bg)(rg-gr)]+ \nonumber \\
&
\bar{b}[(rb-br)(gb-bg)-(gb-bg)(rb-br)] \rbrace\,.
\end{align}

With respect the flavor wave function of fully-charm and -bottom pentaquarks, it is simply expressed as $\chi^{f}_I=QQQ\bar{Q}Q$ $(Q=c,\,b)$, which corresponds to the quark sequence shown in Fig.~\ref{FHPC} and simply gives a total isospin, $I$, equal to zero.

Only $S$-wave fully-heavy pentaquark states shall be considered and thus the total angular momentum, $J$, coincides with the spin, $S$, of the 5-quark system which ranges from $1/2$ to $5/2$. Therefore, we should only focus on the possible spin wave functions of a fully-heavy pentaquark system; the spin wave function of baryon-meson structure of Fig.~\ref{FHPC}(a) can be written as (the third component of spin is taken to be equal to $S$ without loss of generality):
\begin{align}
\label{Spin1}
\chi_{\frac12,\frac12}^{\sigma 1}(5) &= \sqrt{\frac{1}{6}} \chi_{\frac32,-\frac12}^{\sigma}(3) \chi_{11}^{\sigma}
-\sqrt{\frac{1}{3}} \chi_{\frac32,\frac12}^{\sigma}(3) \chi_{10}^{\sigma} \nonumber \\
&
+\sqrt{\frac{1}{2}} \chi_{\frac32,\frac32}^{\sigma}(3) \chi_{1-1}^{\sigma} \\
\chi_{\frac12,\frac12}^{\sigma 2}(5) &= \sqrt{\frac{1}{3}} \chi_{\frac12,\frac12}^{\sigma+}(3) \chi_{10}^{\sigma} -\sqrt{\frac{2}{3}} \chi_{\frac12,-\frac12}^{\sigma+}(3) \chi_{11}^{\sigma} \\
\chi_{\frac12,\frac12}^{\sigma 3}(5) &= \sqrt{\frac{1}{3}} \chi_{\frac12,\frac12}^{\sigma-}(3) \chi_{10}^{\sigma} - \sqrt{\frac{2}{3}} \chi_{\frac12,-\frac12}^{\sigma-}(3) \chi_{11}^{\sigma} \\
\chi_{\frac12,\frac12}^{\sigma 4}(5) &= \chi_{\frac12,\frac12}^{\sigma+}(3) \chi_{00}^{\sigma} \\
\chi_{\frac12,\frac12}^{\sigma 5}(5) &= \chi_{\frac12,\frac12}^{\sigma-}(3) \chi_{00}^{\sigma}
\end{align}
for $S=1/2$, and
\begin{align}
\label{Spin2}
\chi_{\frac32,\frac32}^{\sigma 1}(5) &= \sqrt{\frac{3}{5}}
\chi_{\frac32,\frac32}^{\sigma}(3) \chi_{10}^{\sigma} -\sqrt{\frac{2}{5}} \chi_{\frac32,\frac12}^{\sigma}(3) \chi_{11}^{\sigma} \\
\chi_{\frac32,\frac32}^{\sigma 2}(5) &= \chi_{\frac32,\frac32}^{\sigma}(3) \chi_{00}^{\sigma} \\
\chi_{\frac32,\frac32}^{\sigma 3}(5) &= \chi_{\frac12,\frac12}^{\sigma+}(3) \chi_{11}^{\sigma} \\
\chi_{\frac32,\frac32}^{\sigma 4}(5) &= \chi_{\frac12,\frac12}^{\sigma-}(3) \chi_{11}^{\sigma}
\end{align}
for $S=3/2$, and
\begin{align}
\label{Spin3}
\chi_{\frac52,\frac52}^{\sigma 1}(5) &= \chi_{\frac32,\frac32}^{\sigma}(3) \chi_{11}^{\sigma}
\end{align}
for $S=5/2$.

The possible spin wave functions for the diquark-diquark-antiquark configuration of Fig.~\ref{FHPC}(b), compatible with the quantum numbers that we are investigating, can be summarized as below,
\begin{align}
\label{Spin4}
\chi_{\frac12,\frac12}^{\sigma 6}(5) &= \chi_{00}^{\sigma} \chi_{00}^{\sigma} \chi_{\frac12,\frac12}^{\sigma} \\
\chi_{\frac12,\frac12}^{\sigma 7}(5) &= \sqrt{\frac{2}{3}}\chi_{00}^{\sigma} \chi_{11}^{\sigma} \chi_{\frac12,-\frac12}^{\sigma}-\sqrt{\frac{1}{3}}\chi_{00}^{\sigma} \chi_{10}^{\sigma} \chi_{\frac12,\frac12}^{\sigma} \\
\chi_{\frac12,\frac12}^{\sigma 8}(5) &= \sqrt{\frac{1}{3}}(\chi_{11}^{\sigma} \chi_{1-1}^{\sigma}-\chi_{10}^{\sigma} \chi_{10}^{\sigma}+\chi_{1-1}^{\sigma} \chi_{11}^{\sigma})\chi_{\frac12,\frac12}^{\sigma} \\
\chi_{\frac12,\frac12}^{\sigma 9}(5) &= \sqrt{\frac{1}{3}}(\chi_{11}^{\sigma} \chi_{10}^{\sigma}-\chi_{10}^{\sigma} \chi_{11}^{\sigma})\chi_{\frac12,-\frac12}^{\sigma}- \\ \nonumber
&
\sqrt{\frac{1}{6}}(\chi_{11}^{\sigma} \chi_{1-1}^{\sigma}-\chi_{1-1}^{\sigma} \chi_{11}^{\sigma})\chi_{\frac12,\frac12}^{\sigma}
\end{align}
for $S=1/2$, and
\begin{align}
\label{Spin5}
\chi_{\frac32,\frac32}^{\sigma 5}(5) &= \chi_{00}^{\sigma} \chi_{11}^{\sigma} \chi_{\frac12,\frac12}^{\sigma} \\
\chi_{\frac32,\frac32}^{\sigma 6}(5) &= \sqrt{\frac{1}{2}}(\chi_{11}^{\sigma} \chi_{10}^{\sigma}-\chi_{10}^{\sigma} \chi_{11}^{\sigma})\chi_{\frac12,\frac12}^{\sigma} \\
\chi_{\frac32,\frac32}^{\sigma 7}(5) &= \sqrt{\frac{4}{5}}\chi_{11}^{\sigma} \chi_{11}^{\sigma}\chi_{\frac12,-\frac12}^{\sigma}- \\ \nonumber
&
\sqrt{\frac{1}{10}}(\chi_{11}^{\sigma} \chi_{10}^{\sigma}+\chi_{10}^{\sigma} \chi_{11}^{\sigma})\chi_{\frac12,\frac12}^{\sigma}
\end{align}
for $S=3/2$, and
\begin{align}
\label{Spin6}
\chi_{\frac52,\frac52}^{\sigma 2}(5) &= \chi_{11}^{\sigma} \chi_{11}^{\sigma} \chi_{\frac12,\frac12}^{\sigma}
\end{align}
for $S=5/2$.

Note herein that all these expressions can be easily derived by considering the 3-quark, quark-quark(antiquark) sub-clusters and using SU(2) algebra. Particularly, the spin bases which ranges from Eq.~(\ref{Spin1}) to Eq.~(\ref{Spin6}), have been already derived by us when investigating other multiquark systems, further details can be found in Refs.~\cite{Yang:2015bmv, Yang:2018oqd, gy:2020dcp}.

\subsection{Computational method}

We have already mentioned that two sets of fully-heavy pentaquark configurations, shown in Fig.~\ref{FHPC}, are considered. Consequently, their antisymmetry operators must be categorized. The antisymmetry operator for the baryon-meson structure of Fig.~\ref{FHPC}(a) is 
\begin{equation}
{\cal{A}}_1 = [1-(15)-(25)-(35)][1-(13)-(23)] \,. \label{EE1}
\end{equation}
Meanwhile, the diquark-diquark-antiquark arrangement of Fig.~\ref{FHPC}(b) has an antisymmetry operator
\begin{equation}
{\cal{A}}_2 = 1-(13)-(15)-(23)-(25)+(13)(25) \,. \label{EE2}
\end{equation}

The Rayleigh-Ritz variational principle is employed herein to solve the Schr\"odinger-like 5-body bound state equation because it is one of the most extended tools to solve eigenvalue problems due to its simplicity and flexibility. Given a set of relative motion coordinates, where the center-of-mass kinetic term $T_{CM}$ can be completely eliminated for a non-relativistic system, the $5$-quark system spatial wave function is generally written as:
\begin{align}
\label{eq:WFexp}
&
\psi_{LM_L}(\theta)=[ [ [ \phi_{n_1l_1}({\vec{\rho}} e^{i\theta})\phi_{n_2l_2}({\vec{\lambda}} e^{i\theta})]_{l} \phi_{n_3l_3}({\vec{r}} e^{i\theta}) ]_{l^{\prime}} \nonumber\\
&
\hspace*{1.60cm}  \phi_{n_4l_4}({\vec{R}} e^{i\theta}) ]_{LM_L} \,.
\end{align}
In particular, the four internal Jacobi coordinates of baryon-meson structure of Fig.~\ref{FHPC}(a) are defined in the following way:
\begin{align}
\vec{\rho} &= \vec{x}_1-\vec{x}_2 \,, \\
\vec{\lambda} &= \vec{x}_3 - \frac{{m_1\vec{x}}_1+{m_2\vec{x}}_2}{m_1+m_2} \,,  \\
\vec{r} &= \vec{x}_4-\vec{x}_5 \,, \\
\vec{R} &= \frac{m_1\vec{x}_1+m_2\vec{x}_2 + m_3\vec{x}_3}{m_1+m_2+m_3}-\frac{m_4\vec{x}_4+m_5\vec{x}_5}{m_4+m_5} \,.
\end{align}
And, in the case of diquark-diquark-antiquark configuration of Fig.~\ref{FHPC}(b), they read as
\begin{align}
\vec{\rho} &= \vec{x}_1-\vec{x}_2 \,, \\
\vec{\lambda} &= \vec{x}_3-\vec{x}_5 \,, \\
\vec{r} &= \frac{m_1\vec{x}_1+m_2\vec{x}_2}{m_1+m_2}-\frac{m_3\vec{x}_3+m_5\vec{x}_5}{m_3+m_5} \,,\\
\vec{R} &= \vec{x}_4-\frac{m_1\vec{x}_1+m_2\vec{x}_2 + m_3\vec{x}_3+m_5\vec{x}_5}{m_1+m_2+m_3+m_5}\,.
\end{align}

A crucial point of the Rayleigh-Ritz variational method is the basis expansion of the genuine state's wave function. Herein, the Gaussian expansion method (GEM)~\cite{Hiyama:2003cu} is employed; each relative coordinate is expanded in terms of a Gaussian basis whose sizes are taken in geometric progression. Moreover, as mentioned above, GEM is complemented with the complex scaling method in order to have access at the same time to bound, resonance and scattering states. This approach has been proven before to be quite efficient on solving bound- and resonance-state problem of multiquark systems~\cite{Yang:2015bmv, Yang:2018oqd, Yang:2017rpg, Yang:2021hrb, gy:2020dhts, gy:2020dcp, gy:2020dht}. The interested reader is referred to Ref.~\cite{Yang:2020atz} for a comprehensive summary on the GEM and complex scaling method (CSM) applied to multiquark systems. Consequently, the functional form of the orbital wave functions, $\phi$, shown in Eq.~\eqref{eq:WFexp} is  
\begin{align}
&
\phi_{nlm}(\vec{r}e^{i\theta}\,) = N_{nl} (re^{i\theta})^{l} e^{-\nu_{n} (re^{i\theta})^2} Y_{lm}(\hat{r}) \,.
\end{align}
Only $S$-wave states of fully-heavy pentaquarks are investigated in this work and thus no laborious Racah algebra is needed when computing matrix elements. In other words, the value of the spherical harmonic function $Y_{lm}$ is just a constant, \emph{viz.} $Y_{00}=\sqrt{1/4\pi}$.

Finally, in order to fulfill the Pauli principle, the complete anti-symmetric complex wave-function of 5-quark system is written as
\begin{equation}
\label{TPs}
\Psi_{JM,i,j} (\theta)=\sum_{n=1}^2 {\cal A}_n \left[ \left[ \psi_{L} (\theta) \chi^{\sigma_i}_{S}(5) \right]_{JM} \chi^{f}_I \chi^{c}_j \right] \,,
\end{equation}
where ${\cal A}_n$ is the antisymmetry operator whose expression is either Eq.~(\ref{EE1}) or Eq.~(\ref{EE2}) depending if baryon-meson or diquark-diquark-antiquark are accordingly considered. The use of them is needed because we have constructed an antisymmetric wave function for only two quarks of the 3-quark sub-cluster in baryon-meson configuration or for the quark-quark sub-cluster in diquark-diquark-antiquark arrangement; the remaining quarks in the system are coupled to the wave function by simply considering appropriate Clebsch-Gordan coefficients.


\section{Results}
\label{sec:results}

The $S$-wave low-lying states of fully-charm and -bottom pentaquarks are systematically investigated by taking into account both baryon-meson and diquark-diquark-antiquark configurations. Besides, all possible color structures of each configuration, along with their couplings, are comprehensively considered. The total angular momentum $J$, which coincides with the total spin $S$, ranges from $1/2$ to $5/2$ with a negative parity $P=-1$.

Firstly, real-range investigations are performed on the fully-heavy pentaquarks and Tables~\ref{Gresult1} to~\ref{GresultR6} list our calculated results. In particular, masses of each channel in baryon-meson configuration and diquark-diquark-antiquark one, along with their couplings, are summarized in Tables~\ref{Gresult1},~\ref{Gresult2} and~\ref{Gresult3} for the fully-charm pentaquarks with $J^P=\frac{1}{2}^-$, $\frac{3}{2}^-$ and $\frac{5}{2}^-$, respectively. Besides, similar studies for the fully-bottom pentaquarks are shown in Tables~\ref{Gresult4},~\ref{Gresult5} and~\ref{Gresult6}, respectively. Therein, the two considered configurations of pentaquark system, baryon-meson and diquark-diquark-antiquark, are listed in the first column. They are then assigned an index in the following column. The third one presents a particular combination of spin ($\chi_J^{\sigma_i}$) and color ($\chi_j^c$) wave functions, compatible with the spin-parity quantum numbers considered in each case. The theoretical mass of each channel is shown in the fourth column, and the coupled result for each kind of configuration is presented in the last one. The last row of all those tables indicates the lowest-lying-coupled mass in a complete coupled-channel computation.

In a further step, the CSM is employed only for complete coupled-channel calculations of the $QQQQ\bar{Q}$ systems. Figures~\ref{PP1} to~\ref{PP6} show general distributions of the found complex eigenenergies; possible resonance states are indicated inside (orange) circles. Meanwhile, Tables~\ref{GresultR1},~\ref{GresultR2} and~\ref{GresultR3} are about compositeness of the obtained fully-charm resonances within $\frac{1}{2}^-$, $\frac{3}{2}^-$ and $\frac{5}{2}^-$, respectively. The resonance pole, encoded by its mass and width $(M+i\Gamma)$, is listed in the first column; its inner structure elucidated by quark-quark and quark-antiquark distances is shown in the second column; and the last column presents the dominant component of each resonance. The corresponding analyses for fully-bottom pentaquarks are collected in Tables~\ref{GresultR4},~\ref{GresultR5} and~\ref{GresultR6}, respectively.

We proceed now to describe in detail our theoretical findings.


\subsection{Fully-charm pentaquarks}

{\bf The $\bm{J^P=\frac12^-}$ sector:} Table~\ref{Gresult1} shows our results in this case. There is only one color-singlet baryon-meson channel, $\Omega_{ccc}J/\psi$. Its calculated lowest mass is just the non-interacting baryon-meson threshold value 7899 MeV. Meanwhile, the five hidden-color configurations of $\Omega_{ccc}J/\psi$ are above 8.0 GeV, ranging from $8.08$ to $8.56$ GeV. Three hidden-color diquark-diquark-antiquark channels are around $8.1$ GeV. When a coupled-channel calculation is performed in each exotic configuration: hidden-color baryon-meson and diquark-diquark-antiquark, their lowest masses are $8.02$ and $8.04$ GeV, respectively. Therefore, no bound state is found and this fact remains unchanged within a complete coupled-channel real-range calculation.

\begin{table}[!t]
\caption{\label{Gresult1} Lowest-lying fully-charm pentaquark states with $J^P=1/2^-$ calculated within a real range formulation of the potential quark model.
Baryon-meson and diquark-diquark-antiquark configurations are listed in the first column, the superscripts 1 and 8 stand for color-singlet and -octet state, respectively. Each channel is assigned an index in the 2nd column, it reflects a particular combination of spin ($\chi_J^{\sigma_i}$) and color ($\chi_j^c$) wave functions that are shown explicitly in the 3rd column. The theoretical mass obtained in each channel is shown in the 4th column and the coupled result for each kind of configuration is presented in the last column.
When a complete coupled-channel calculation is performed, last row of the table indicates the lowest-lying mass.
(unit: MeV).}
\begin{ruledtabular}
\begin{tabular}{lcccc}
~~Channel   & Index & $\chi_J^{\sigma_i}$;~$\chi_j^c$ & $M$ & Mixed~~ \\
        &   &$[i; ~j]$ & & \\
\hline
$(\Omega_{ccc} J/\psi)^1$          & 1  & [1;~1]  & $7899$ & $7899$ \\[2ex]
$(\Omega_{ccc} J/\psi)^8 $        & 2  & [5;~2]   & $8088$ &  \\
$(\Omega_{ccc} J/\psi)^8 $        & 3  & [4;~3]  & $8082$ & \\
$(\Omega_{ccc} J/\psi)^8 $        & 4  & [3;~2]   & $8145$ & \\
$(\Omega_{ccc} J/\psi)^8 $        & 5  & [2;~3]  & $8144$ & \\
$(\Omega_{ccc} J/\psi)^8 $        & 6  & [1;~3]   & $8556$ & $8022$ \\ [2ex]
$(cc)(cc)^*\bar{c}$          & 7  & [7;~4]  & $8142$ & \\
$(cc)^*(cc)^*\bar{c}$  & 8  & [9;~5]   & $8098$ &  \\
$(cc)^*(cc)^*\bar{c}$      & 9   & [8;~5]  & $8167$ & $8045$ \\[2ex]
\multicolumn{4}{c}{Complete coupled-channel:} & $7899$
\end{tabular}
\end{ruledtabular}
\end{table}

\begin{figure}[!t]
\epsfxsize=3.8in \epsfbox{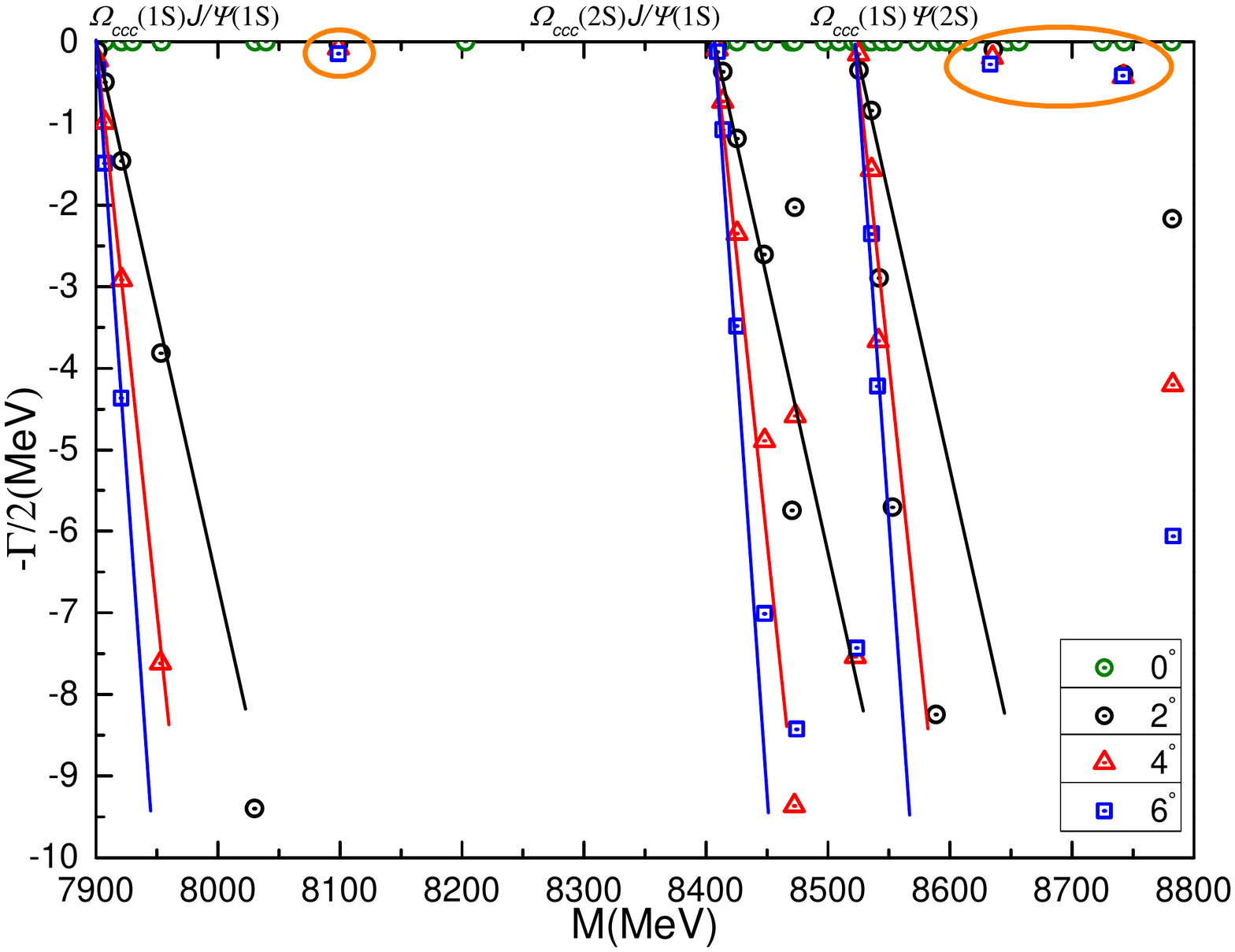}
\caption{Complex eigenenergies of coupled-channel calculation for various $\theta$ within $J^P=\frac12^-$.} \label{PP1}
\end{figure}

In a further step where the complex scaling method is employed in a fully coupled-channel calculation, the scattering states of $\Omega_{ccc}(1S)J/\psi(1S)$, $\Omega_{ccc}(2S)J/\psi(1S)$ and $\Omega_{ccc}(1S)J/\psi(2S)$ are well shown in Fig.~\ref{PP1} within an energy gap $7.9-8.8$ GeV. Therein, with a rotated angle varied from $0^\circ$ to $6^\circ$, the vast majority of complex energies are aligned well on their corresponding threshold lines. However, three fixed poles, which are independent of $\theta$, are clearly obtained. They are quite narrow resonances, and the complex energies, denoted as $M+i\Gamma$, are $(8098+i0.16)\,\text{MeV}$, $(8634+i0.38)\,\text{MeV}$ and $(8742+i0.86)\,\text{MeV}$, respectively.

\begin{table}[!t]
\caption{\label{GresultR1} Compositeness of the exotic resonances for the $J^P=1/2^-$ fully-charm pentaquarks, obtained when a complete coupled-channel analysis using CSM is performed. Particularly, the first column is the resonance pole labeled by $M+i\Gamma$, unit in MeV; the second one is the distance between any two quarks or quark-antiquark, unit in fm; and the last column is the component of each resonance state (S: baryon-meson structure in color-singlet channel; H: baryon-meson structure in hidden-color channel; Di: diquark-diquark-antiquark configuration).}
\begin{ruledtabular}
\begin{tabular}{lccc}
Resonance   & $(r_{cc},\,r_{c\bar{c}})$ & \multicolumn{2}{c}{Component} \\
\hline
$8098+i0.16$     & $(0.53,\,0.56)$   & \multicolumn{2}{c}{Di: 100\%} \\
$8634+i0.38$    & $(0.75,\,0.64)$   & \multicolumn{2}{c}{Di: 100\%}  \\
$8742+i0.86$    & $(0.76,\,0.71)$   & \multicolumn{2}{c}{S: 46\%; H: 52.7\%; Di: 1.3\%}  \\
\end{tabular}
\end{ruledtabular}
\end{table}

We compute the quark--(anti-)quark distances and the dominant wavefunction component in our analysis about the compositeness of these three resonance poles. Table~\ref{GresultR1} shows our results, one can conclude that these 5-quark exotic resonances are compact with a size around $0.5-0.8$ fm. This nature is also confirmed by their dominant component which is either diquark-diquark-antiquark or hidden-color baryon-meson. It is worth noting that our result on the first resonance at $(8098+i0.16)$~MeV is also supported by Ref.~\cite{Wang:2021xao}, whose calculated mass upper limit for the fully-charm pentaquark of diquark-diquark-antiquark nature and $J^P=\frac{1}{2}^-$ quantum numbers is $8.08$ GeV.

{\bf The $\bm{J^P=\frac32^-}$ sector:} Table~\ref{Gresult2} shows that there are also nine channels to be considered in this fully-charm pentaquark sector. In particular, there are two color-singlet baryon-meson structures, $\Omega_{ccc}\eta_c$ and $\Omega_{ccc}J/\psi$, four hidden-color channels of the same baryon-meson nature and three diquark-diquark-antiquark cases. First of all, when a calculation in each single channel is performed, the lowest mass $7765$ MeV is just the theoretical threshold value of $\Omega_{ccc}\eta_c$; moreover, there is another calculated color-singlet mass with a value of $7899$ MeV which is located at the $\Omega_{ccc}J/\psi$ threshold; and the other exotic channels are generally lying within an interval of $7.99-8.26$ GeV. When a coupled-channel computation is performed in each particular pentaquark configuration, the masses are all located above $\Omega_{ccc}\eta_c$ threshold, and they are $7765$ MeV, $7982$ MeV and $8095$ MeV for the color-singlet baryon-meson, hidden-color baryon-meson and diquark-diquark-antiquark cases, respectively. The last row of Table~\ref{Gresult2} shows the lowest computed mass, $7765$ MeV, when a complete coupled-channel computation in real range formalism is performed; therefore, no bound states are found for fully-charm pentaquarks with $J^P=3/2^-$ quantum numbers.

\begin{table}[!t]
\caption{\label{Gresult2} Lowest-lying fully-charm pentaquark states with $J^P=3/2^-$ calculated within a real range formulation of the potential quark model. The results are similarly organized as those in Table~\ref{Gresult1}.
(unit: MeV).}
\begin{ruledtabular}
\begin{tabular}{lcccc}
~~Channel   & Index & $\chi_J^{\sigma_i}$;~$\chi_j^c$ & $M$ & Mixed~~ \\
        &   &$[i; ~j]$ & & \\
\hline
$(\Omega_{ccc} \eta_c)^1$          & 1  & [2;~1]  & $7765$ & \\
$(\Omega_{ccc} J/\psi)^1 $        & 2  & [1;~1]   & $7899$ & $7765$ \\[2ex]
$(\Omega_{ccc} J/\psi)^8 $        & 3  & [4;~2]  & $7992$ & \\
$(\Omega_{ccc} J/\psi)^8 $        & 4  & [3;~3]   & $7987$ & \\
$(\Omega_{ccc} \eta_c)^8 $        & 5  & [2;~3]  & $8155$ & \\
$(\Omega_{ccc} J/\psi)^8 $        & 6  & [1;~3]   & $8259$ & $7982$ \\ [2ex]
$(cc)(cc)^*\bar{c}$          & 7  & [5;~4]  & $8104$ & \\
$(cc)^*(cc)^*\bar{c}$  & 8  & [6;~5]   & $8131$ &  \\
$(cc)^*(cc)^*\bar{c}$      & 9   & [7;~5]  & $8095$ & $8095$ \\[2ex]
\multicolumn{4}{c}{Complete coupled-channel:} & $7765$
\end{tabular}
\end{ruledtabular}
\end{table}

\begin{figure}[ht]
\epsfxsize=3.8in \epsfbox{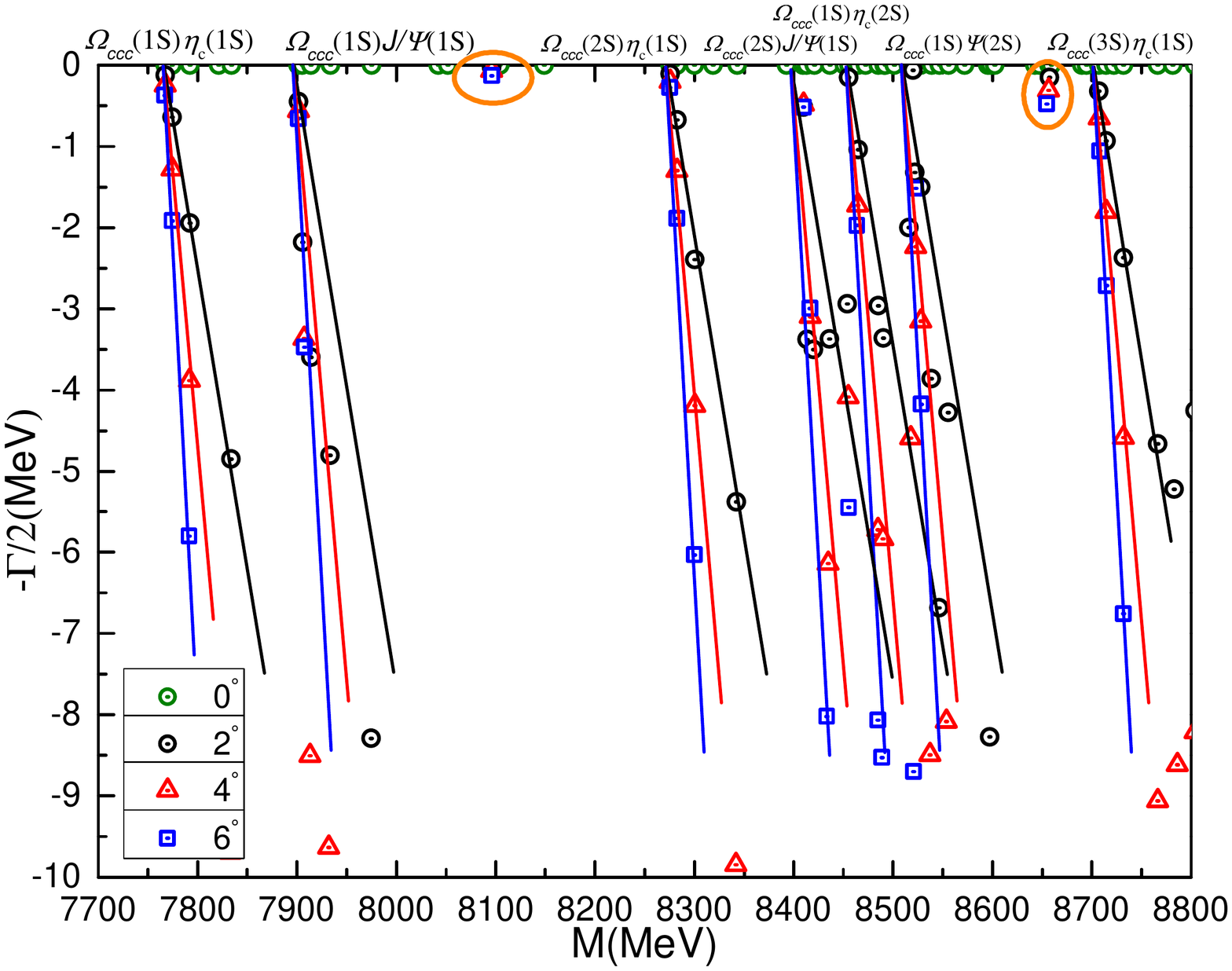}
\caption{Complex eigenenergies of coupled-channel calculation for various $\theta$ within $J^P=\frac32^-$.} \label{PP2}
\end{figure}

Figure~\ref{PP2} presents the distribution of complex eigenenergies of $J^P=3/2^-$ fully-charm pentaquark system when a complete coupled-channel analysis is carried out using the CSM. Within the energy interval $7.7-8.8$ GeV, the ground states of $\Omega_{ccc}\eta_c$ and $\Omega_{ccc}J/\psi$, along with their radial excitations, are well presented. These baryon-meson states are of scattering nature, with calculated discrete energies always going down with respect to the variation of complex angle $\theta$. Nevertheless, two extremely narrow resonances survive among these continuum states.

\begin{table}[!t]
\caption{\label{GresultR2} Compositeness of the exotic resonances for the $J^P=3/2^-$ fully-charm pentaquarks, obtained when a complete coupled-channel analysis using CSM is performed. The results are similarly organized as those in Table~\ref{GresultR1}.}
\begin{ruledtabular}
\begin{tabular}{lccc}
Resonance   & $(r_{cc},\,r_{c\bar{c}})$ & \multicolumn{2}{c}{Component} \\
\hline
$8095+i0.14$     & $(0.53,\,0.55)$   & \multicolumn{2}{c}{Di: 100\%} \\
$8656+i0.62$    & $(0.76,\,0.65)$   & \multicolumn{2}{c}{Di: 100\%}  \\
\end{tabular}
\end{ruledtabular}
\end{table}

The theoretical poles are located at $(8095+i0.14)\,\text{MeV}$ and $(8656+i0.62)\,\text{MeV}$, respectively. Hence, it seems that the lowest resonances of fully-charm pentaquarks with $\frac{1}{2}^-$ and $\frac{3}{2}^-$ quantum numbers are degenerate. Additionally, Table~\ref{GresultR2} analyzes the compositeness of the two poles found in this channel and one may conclude that they are quite compact whose sizes are less than $0.8$~fm and their dominant wavefunction's components are of diquark-diquark-antiquark type.

{\bf The $\bm{J^P=\frac52^-}$ sector:} In the highest spin case of fully-charm pentaquarks there are only three possible configurations under consideration, \emph{i.e.} just one state of singlet-color baryon-meson, hidden-color baryon-meson and diquark-diquark-antiquark. The calculated mass results are listed in Table~\ref{Gresult3}; it is worth highlighting that the unbound nature of the $\Omega_{ccc}J/\psi$, whose theoretical mass is 7899~MeV, remains unchanged whether coupling effects are included or not. Besides, masses of the other two exotic structures are 8421~MeV and 8137~MeV, respectively.

\begin{table}[!t]
\caption{\label{Gresult3} Lowest-lying fully-charm pentaquark states with $J^P=5/2^-$ calculated within a real range formulation of the potential quark model. The results are similarly organized as those in Table~\ref{Gresult1}.
(unit: MeV).}
\begin{ruledtabular}
\begin{tabular}{lcccc}
~~Channel   & Index & $\chi_J^{\sigma_i}$;~$\chi_j^c$ & $M$ & Mixed~~ \\
        &   &$[i; ~j]$ &  & \\
\hline
$(\Omega_{ccc} J/\psi)^1 $        & 1  & [1;~1]   & $7899$ & $7899$ \\[2ex]
$(\Omega_{ccc} J/\psi)^8 $       & 2  & [1;~3]   & $8421$ & $8421$ \\ [2ex]
$(cc)^*(cc)^*\bar{c}$      & 3   & [1;~5]  & $8137$ & $8137$ \\[2ex]
\multicolumn{4}{c}{Complete coupled-channel:} & $7899$
\end{tabular}
\end{ruledtabular}
\end{table}

\begin{figure}[ht]
\epsfxsize=3.8in \epsfbox{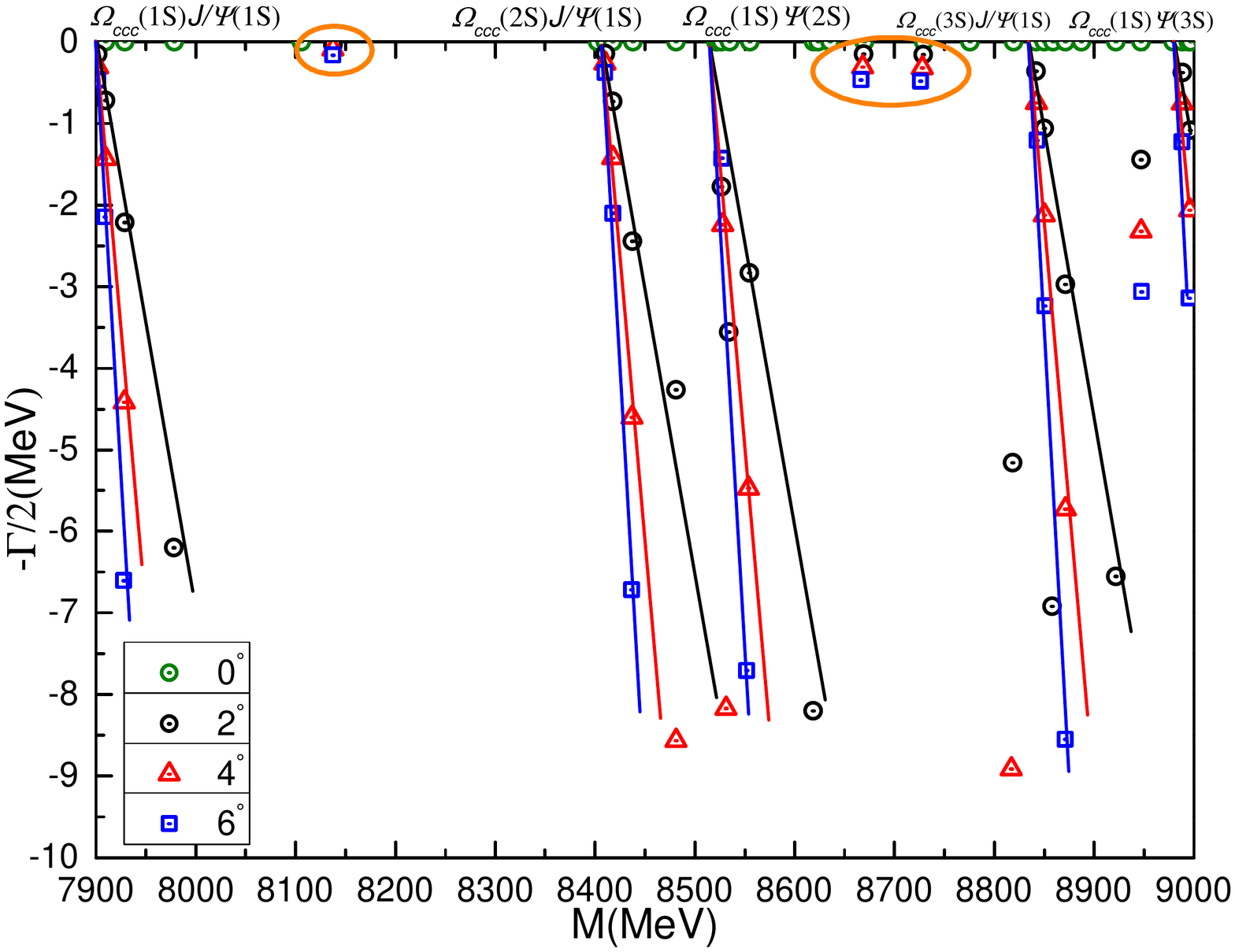}
\caption{Complex eigenenergies of coupled-channel calculation for various $\theta$ within $J^P=\frac52^-$.} \label{PP3}
\end{figure}

Similarly to the previous $J^P$ quantum quantum numbers, narrow resonance states are available in a fully coupled-channel investigation using the complex scaling method. Figure~\ref{PP3} shows our results, the scattering states of $\Omega_{ccc}(1S)J/\psi(1S)$ and its radial excitations are clearly presented within the energy range $7.9-9.0$ GeV; meanwhile, three stable resonances are circled therein. Their complex energies are $(8137+i0.18)\,\text{MeV}$, $(8668+i0.62)\,\text{MeV}$ and $(8727+i0.64)\,\text{MeV}$. All of them are still compact multiquarks whose size and exotic nature is indicated in Table~\ref{GresultR3}.

\begin{table}[!t]
\caption{\label{GresultR3} CCompositeness of the exotic resonances for the $J^P=5/2^-$ fully-charm pentaquarks, obtained when a complete coupled-channel analysis using CSM is performed. The results are similarly organized as those in Table~\ref{GresultR1}.}
\begin{ruledtabular}
\begin{tabular}{lccc}
Resonance   & $(r_{cc},\,r_{c\bar{c}})$ & \multicolumn{2}{c}{Component} \\
\hline
$8137+i0.18$     & $(0.53,\,0.57)$   & \multicolumn{2}{c}{Di: 100\%} \\
$8668+i0.62$    & $(0.74,\,0.68)$   & \multicolumn{2}{c}{Di: 100\%}  \\
$8727+i0.64$    & $(0.75,\,0.65)$   & \multicolumn{2}{c}{S: 11.1\%; H: 37.9\%; Di: 51\%}  \\
\end{tabular}
\end{ruledtabular}
\end{table}

\subsection{Fully-bottom pentaquarks}

{\bf The $\bm{J^P=\frac12^-}$ sector:} Table~\ref{Gresult4} collects the theoretical masses obtained in each fully-bottom pentaquark configuration and also taking into account their couplings within the same arrangement. In particular, the color-singlet $\Omega_{bbb}\Upsilon$ configuration has a mass of 23821~MeV, the hidden-color cases of the same arrangement are located at around 24.03~GeV, except for one higher state with mass at 24.44~GeV, and when performing coupled-channel calculation the mass is 23.96~GeV, which is above the lowest baryon-meson threshold. As for the three diquark-diquark-antiquark channels, their masses are around 24.06~GeV and the coupling is also weak with a coupled-mass generated at 24.04~GeV. At last, when a complete coupled-channel study is considered in real range approximation, the nature of the scattering state $\Omega_{bbb}\Upsilon$ is not unchanged.

\begin{table}[!t]
\caption{\label{Gresult4}  Lowest-lying fully-bottom pentaquark states with $J^P=1/2^-$ calculated within a real range formulation of the potential quark model. The results are similarly organized as those in Table~\ref{Gresult1}.
(unit: MeV).}
\begin{ruledtabular}
\begin{tabular}{lcccc}
~~Channel   & Index & $\chi_J^{\sigma_i}$;~$\chi_j^c$ & $M$ & Mixed~~ \\
        &   &$[i; ~j]$ & & \\
\hline
$(\Omega_{bbb} \Upsilon)^1$          & 1  & [1;~1]  & $23821$ & $23821$ \\[2ex]
$(\Omega_{bbb} \Upsilon)^8 $        & 2  & [5;~2]   & $24028$ &  \\
$(\Omega_{bbb} \Upsilon)^8 $        & 3  & [4;~3]  & $24016$ & \\
$(\Omega_{bbb} \Upsilon)^8 $        & 4  & [3;~2]   & $24032$ & \\
$(\Omega_{bbb} \Upsilon)^8 $        & 5  & [2;~3]  & $24075$ & \\
$(\Omega_{bbb} \Upsilon)^8 $        & 6  & [1;~3]   & $24443$ & $23959$ \\ [2ex]
$(bb)(bb)^*\bar{b}$          & 7  & [7;~4]  & $24066$ & \\
$(bb)^*(bb)^*\bar{b}$  & 8  & [9;~5]   & $24055$ &  \\
$(bb)^*(bb)^*\bar{b}$      & 9   & [8;~5]  & $24093$ & $24035$ \\[2ex]
\multicolumn{4}{c}{Complete coupled-channel:} & $23821$
\end{tabular}
\end{ruledtabular}
\end{table}

\begin{figure}[ht]
\epsfxsize=3.8in \epsfbox{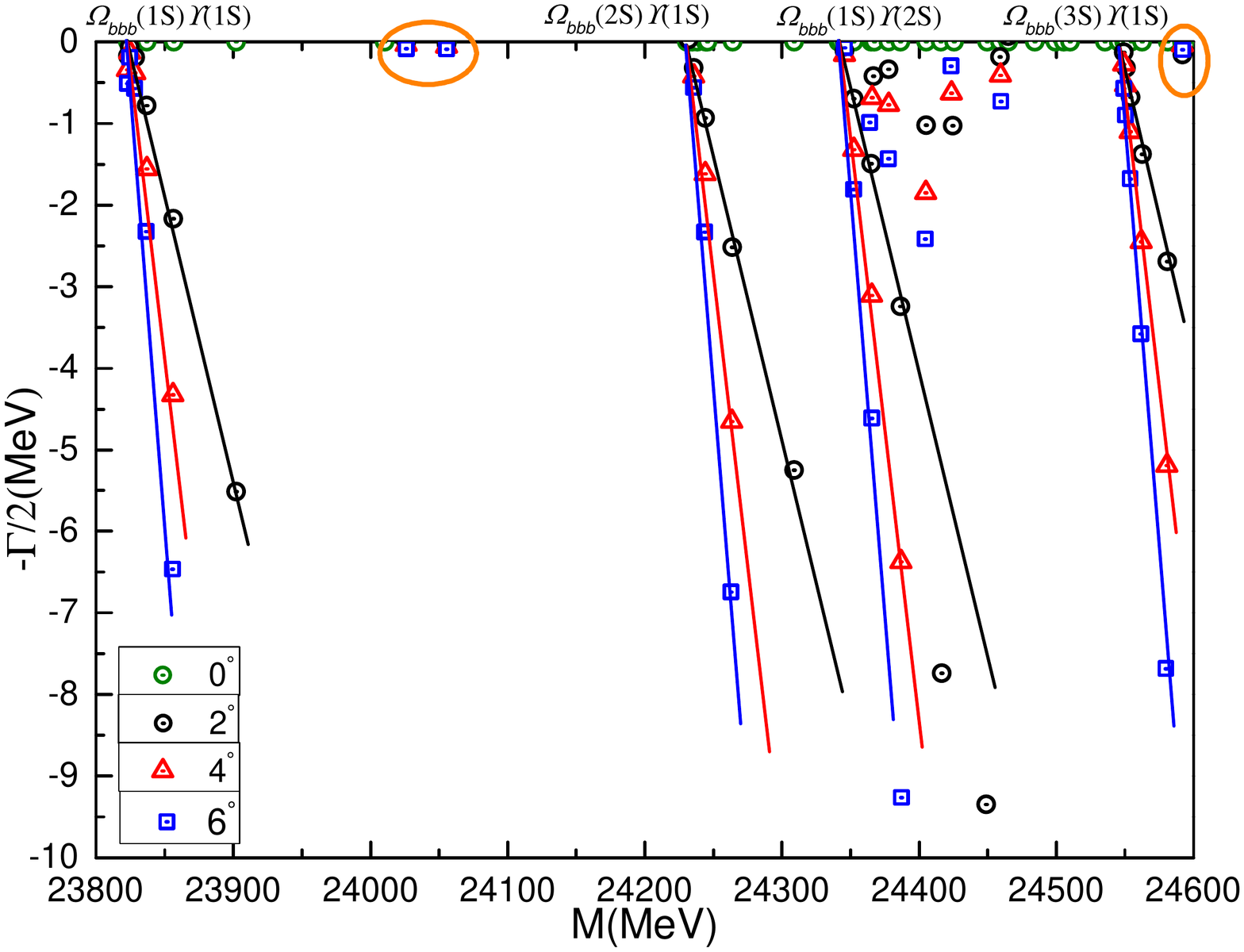}
\caption{Complex eigenenergies of coupled-channel calculation for various $\theta$ within $J^P=\frac12^-$.} \label{PP4}
\end{figure}

Some interesting findings are obtained when a fully coupled-channel investigation has been extended to the complex range. Figure~\ref{PP4} shows the distribution of calculated eigenenergies within the interval 23.8-24.6~GeV. Therein, the scattering nature of $\Omega_{bbb}\Upsilon$ in both ground and radial excitation states are clearly captured. However, three fixed poles are obtained as the complex angle is rotated and their energies read (24026+i0.06)~MeV, (24055+i0.12)~MeV and (24591+i0.08)~MeV. These resonances are extremely narrow and, from Table~\ref{GresultR4}, their sizes are close to 0.3~fm. From the same table, the first two resonances mostly consist of exotic color structures $(\sim100\%)$, and there is a strong coupling between color-singlet $(46.2\%)$ and exotic color $(53.8\%)$ channels for the third one. Finally, the diquark-diquark-antiquark resonance with eigenenergy (24055+i0.12)~MeV is compatible with the one obtained in Ref.~\cite{Wang:2021xao}.

\begin{table}[!t]
\caption{\label{GresultR4} Compositeness of the exotic resonances for the $J^P=1/2^-$ fully-bottom pentaquarks, obtained when a complete coupled-channel analysis using CSM is performed. The results are similarly organized as those in Table~\ref{GresultR1}.}
\begin{ruledtabular}
\begin{tabular}{lccc}
Resonance   & $(r_{bb},\,r_{b\bar{b}})$ & \multicolumn{2}{c}{Component} \\
\hline
$24026+i0.06$     & $(0.29,\,0.30)$   & \multicolumn{2}{c}{S: 4.6\%; H: 36.5\%; Di: 58.9\%} \\
$24055+i0.12$    & $(0.29,\,0.31)$   & \multicolumn{2}{c}{Di: 100\%}  \\
$24591+i0.08$    & $(0.44,\,0.42)$   & \multicolumn{2}{c}{S: 46.2\%; H: 52.3\%; Di: 1.5\%}  \\
\end{tabular}
\end{ruledtabular}
\end{table}

{\bf The $\bm{J^P=\frac32^-}$ sector:} Two color-singlet baryon-meson, four hidden-color baryon-meson and three diquark-diquark-antiquark channels are considered in this case. Table~\ref{Gresult5} shows that the mass of the color-singlet $\Omega_{bbb}\eta_b$ and $\Omega_{bbb}\Upsilon$ are 23.76~GeV and 23.82~GeV, respectively, indicating their scattering nature. Masses of the four hidden-color channels are slightly higher, lying in an energy interval from 23.94 GeV to 24.37 GeV. Besides, the three diquark-diquark-antiquark channels are generally located at $\sim$24.05~GeV. When computing in coupled-channels mode using real range formulation, two exotic color structures reduce their masses in about 20~MeV and this fact does not hold for color-singlet channels. In any case, the lowest mass remains located at the $\Omega_{bbb}\eta_b$ threshold theoretical value, $23.76~GeV$, even if a complete coupled-channel computation is performed.

\begin{table}[!t]
\caption{\label{Gresult5} Lowest-lying fully-bottom pentaquark states with $J^P=3/2^-$ calculated within a real range formulation of the potential quark model. The results are similarly organized as those in Table~\ref{Gresult1}.
(unit: MeV).}
\begin{ruledtabular}
\begin{tabular}{lcccc}
~~Channel   & Index & $\chi_J^{\sigma_i}$;~$\chi_j^c$ & $M$ & Mixed~~ \\
        &   &$[i; ~j]$ & & \\
\hline
$(\Omega_{bbb} \eta_b)^1$          & 1  & [2;~1]  & $23759$ & \\
$(\Omega_{bbb} \Upsilon)^1 $        & 2  & [1;~1]   & $23821$ & $23759$ \\[2ex]
$(\Omega_{bbb} \Upsilon)^8 $        & 3  & [4;~2]  & $23955$ & \\
$(\Omega_{bbb} \Upsilon)^8 $        & 4  & [3;~3]   & $23942$ & \\
$(\Omega_{bbb} \eta_b)^8 $        & 5  & [2;~3]  & $24347$ & \\
$(\Omega_{bbb} \Upsilon)^8 $        & 6  & [1;~3]   & $24374$ & $23936$ \\ [2ex]
$(bb)(bb)^*\bar{b}$          & 7  & [5;~4]  & $24045$ & \\
$(bb)^*(bb)^*\bar{b}$  & 8  & [6;~5]   & $24074$ &  \\
$(bb)^*(bb)^*\bar{b}$      & 9   & [7;~5]  & $24054$ & $24035$ \\[2ex]
\multicolumn{4}{c}{Complete coupled-channel:} & $23759$
\end{tabular}
\end{ruledtabular}
\end{table}

\begin{figure}[ht]
\epsfxsize=3.8in \epsfbox{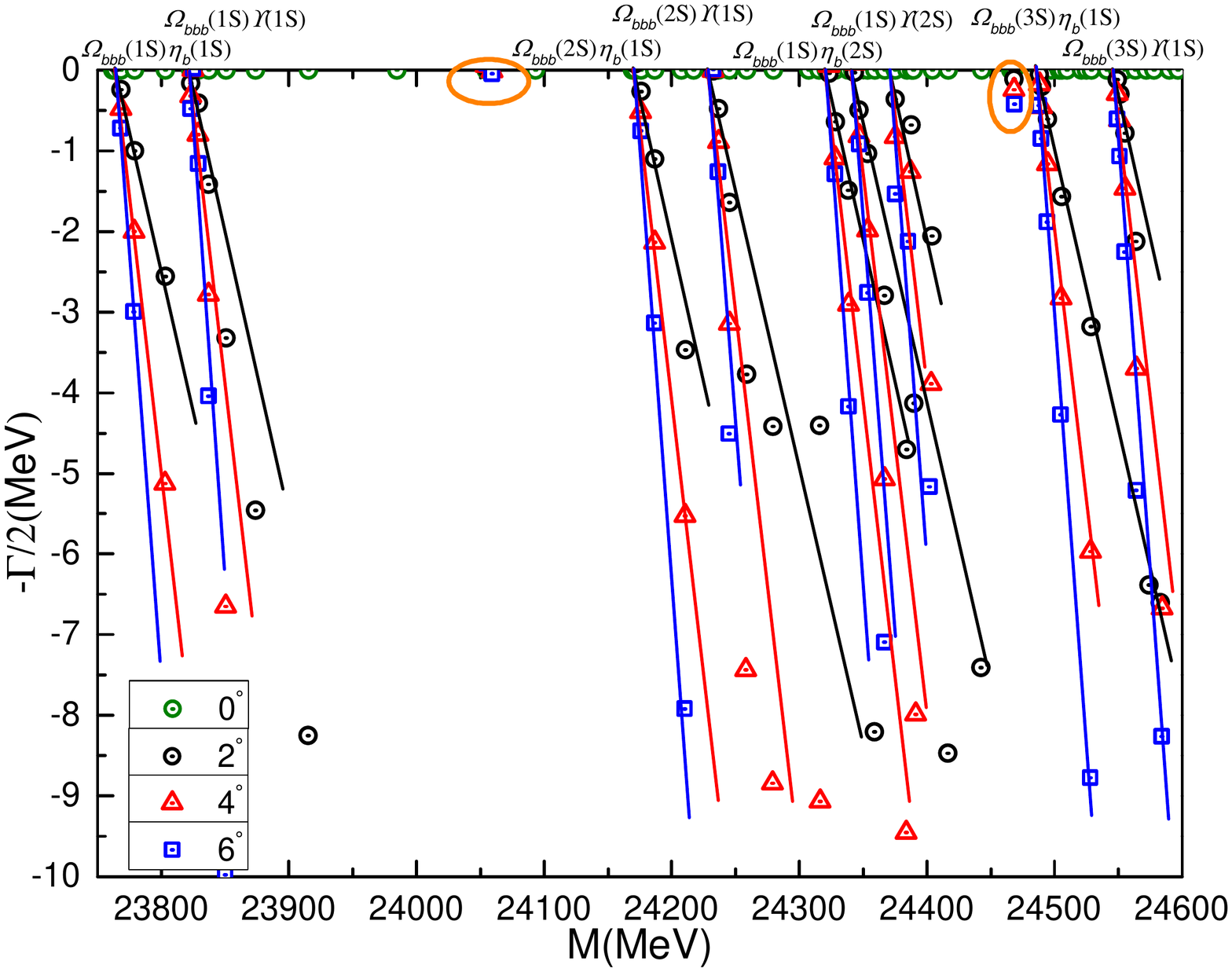}
\caption{Complex eigenenergies of coupled-channel calculation for various $\theta$ within $J^P=\frac32^-$.} \label{PP5}
\end{figure}

In a fully coupled-channel investigation in which the CSM is used, two narrow resonances are obtained and shown in Fig.~\ref{PP5}. Their complex energies read (24059+i0.02)~MeV and (24468+i0.22)~MeV. Furthermore, by analyzing their compositeness in Table~\ref{GresultR5}, both compact and exotic color features are obviously shown. Finally, note that the 23.7-24.6~GeV energy region is plagued of discrete poles belonging to the ground and radial excitation scattering states of $\Omega_{bbb}\eta_b$ and $\Omega_{bbb}\Upsilon$.

\begin{table}[!t]
\caption{\label{GresultR5} Compositeness of the exotic resonances for the $J^P=3/2^-$ fully-bottom pentaquarks, obtained when a complete coupled-channel analysis using CSM is performed. The results are similarly organized as those in Table~\ref{GresultR1}.}
\begin{ruledtabular}
\begin{tabular}{lccc}
Resonance   & $(r_{bb},\,r_{b\bar{b}})$ & \multicolumn{2}{c}{Component} \\
\hline
$24059+i0.02$     & $(0.30,\,0.31)$   & \multicolumn{2}{c}{S: 4.5\%; H: 39.3\%; Di: 56.2\%} \\
$24468+i0.22$    & $(0.34,\,0.49)$   & \multicolumn{2}{c}{Di: 100\%}  \\
\end{tabular}
\end{ruledtabular}
\end{table}

{\bf $\bm{J^P=\frac52^-}$ sector:} There are three channels for the highest spin sector of the fully-bottom pentaquarks. Our calculated results are listed in Table~\ref{Gresult6}. Firstly, a bound state is not possible here with the lowest mass of the color-singlet $\Omega_{bbb}\Upsilon$ channel equal to its theoretical threshold 23.82~GeV. This value remains unchanged in a fully coupled-channel case. Meanwhile, the hidden-color baryon-meson and diquark-diquark-antiquark channels are both excited, and their masses are 24.30~GeV and 24.08~GeV, respectively.

\begin{table}[!t]
\caption{\label{Gresult6} Lowest-lying fully-bottom pentaquark states with $J^P=5/2^-$ calculated within a real range formulation of the potential quark model. The results are similarly organized as those in Table~\ref{Gresult1}.
(unit: MeV).}
\begin{ruledtabular}
\begin{tabular}{lcccc}
~~Channel   & Index & $\chi_J^{\sigma_i}$;~$\chi_j^c$ & $M$ & Mixed~~ \\
        &   &$[i; ~j]$ & & \\
\hline
$(\Omega_{bbb} \Upsilon)^1 $        & 1  & [1;~1]   & $23821$ & $23821$ \\[2ex]
$(\Omega_{bbb} \Upsilon)^8 $       & 2  & [1;~3]   & $24302$ & $24302$ \\ [2ex]
$(bb)^*(bb)^*\bar{b}$      & 3   & [1;~5]  & $24076$ & $24076$ \\[2ex]
\multicolumn{4}{c}{Complete coupled-channel:} & $23821$
\end{tabular}
\end{ruledtabular}
\end{table}

\begin{figure}[ht]
\epsfxsize=3.8in \epsfbox{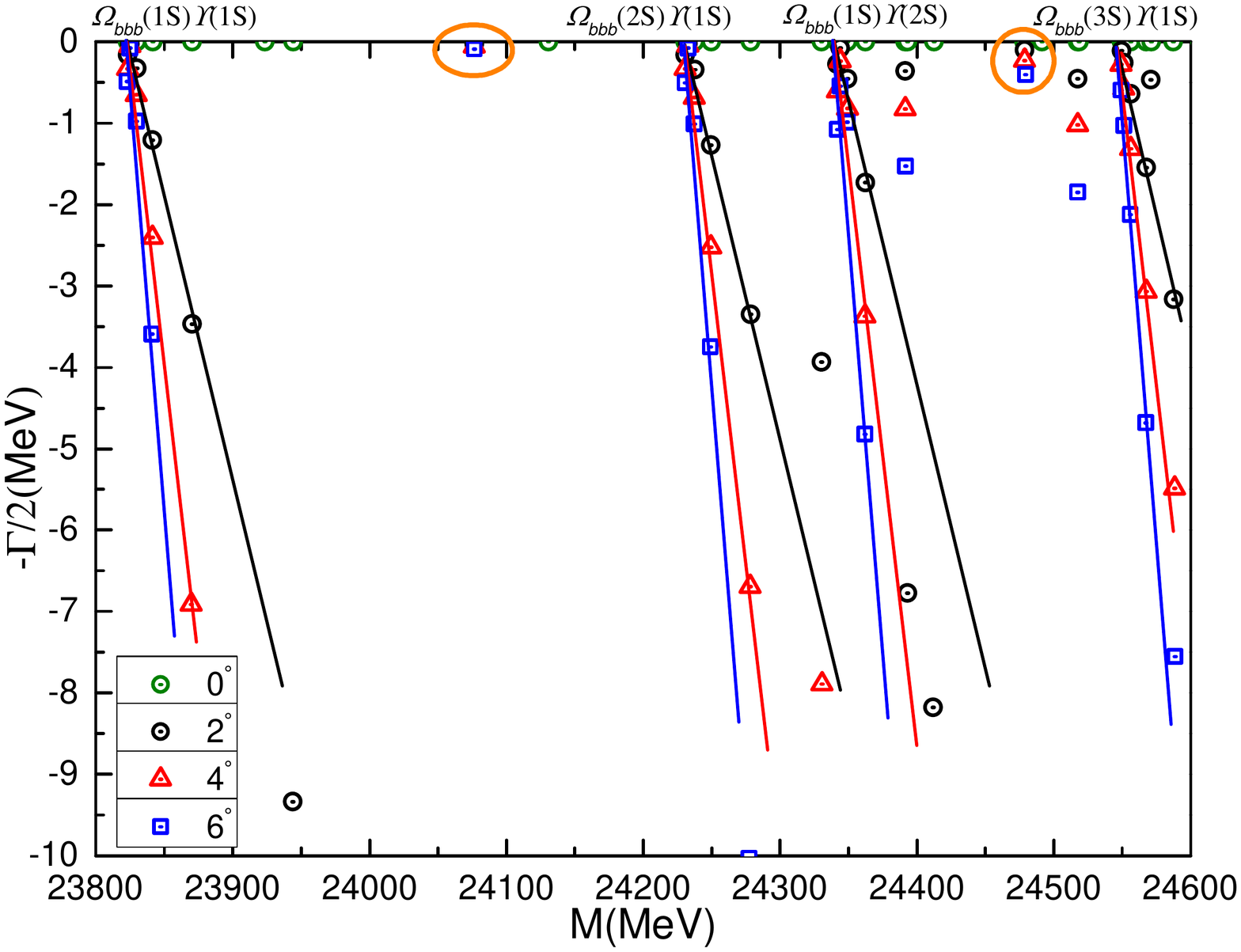}
\caption{Complex eigenenergies of coupled-channel calculation for various $\theta$ within $J^P=\frac52^-$.} \label{PP6}
\end{figure}

In a further calculation, where the CSM is employed in a complete coupled-channel study, some narrow resonances are obtained. Within the energy region 23.8-24.6~GeV of Fig.~\ref{PP6}, two stable poles are circled among the scattering states of $\Omega_{bbb}\Upsilon$. Details on their nature are listed in Table~\ref{GresultR6}. In particular, these two resonances have masses (24076+i0.12)~MeV and (24478+i0.46)~MeV, and both are compact diquark-diquark-antiquark configuration.

\begin{table}[!t]
\caption{\label{GresultR6} Compositeness of the exotic resonances for the $J^P=5/2^-$ fully-bottom pentaquarks, obtained when a complete coupled-channel analysis using CSM is performed. The results are similarly organized as those in Table~\ref{GresultR1}.}
\begin{ruledtabular}
\begin{tabular}{lccc}
Resonance   & $(r_{bb},\,r_{b\bar{b}})$ & \multicolumn{2}{c}{Component} \\
\hline
$24076+i0.12$    & $(0.29,\,0.31)$   & \multicolumn{2}{c}{Di: 100\%} \\
$24478+i0.46$    & $(0.32,\,0.49)$   & \multicolumn{2}{c}{Di: 100\%}  \\
\end{tabular}
\end{ruledtabular}
\end{table}

\begin{table}[!t]
\caption{\label{GresultCCT} Summary of the resonances found in the fully-charm and -bottom pentaquarks. The first column shows the spin-parity of each singularity, the second column refers to the dominant channel (S: color-singlet channel, H: hidden-color channel, Di: diquark-diquark-antiquark channel), the last column shows the corresponding complex eigenenergy, $E=M+i\Gamma$. (unit: MeV).}
\begin{ruledtabular}
\begin{tabular}{llc}
\multicolumn{3}{c}{Fully-charm pentaquarks}\\
~ $J^P$ & Dominant Channel   & Complex Energy~~ \\
\hline
~~$\frac{1}{2}^-$  & Di(100\%)   & $8098+i0.16$~~  \\
            & Di(100\%)   & $8634+i0.38$~~ \\
            & S(46\%)+H(53\%)   & $8742+i0.86$~~ \\[2ex]
~~$\frac{3}{2}^-$  & Di(100\%)   & $8095+i0.14$~~ \\
            & Di(100\%)   & $8656+i0.62$~~ \\[2ex]
~~$\frac{5}{2}^-$  & Di(100\%)   & $8137+i0.18$~~ \\
                  & Di(100\%)   & $8668+i0.62$~~ \\
                  & H(38\%)+Di(51\%)    & $8727+i0.64$~~ \\
\hline
\hline
\multicolumn{3}{c}{Fully-bottom pentaquarks}\\
~ $J^P$ & Dominant Channel   & Complex Energy~~ \\
\hline
~~$\frac{1}{2}^-$  & H(37\%)+Di(59\%)  & $24026+i0.06$~~  \\
            & Di(100\%)   & $24055+i0.12$~~ \\
            & S(46\%)+H(52\%)   & $24591+i0.08$~~ \\[2ex]
~~$\frac{3}{2}^-$  & H(39\%)+Di(56\%)    & $24059+i0.02$~~ \\
            & Di(100\%)   & $24468+i0.22$~~ \\[2ex]
~~$\frac{5}{2}^-$  & Di(100\%)    & $24076+i0.12$~~ \\
                  & Di(100\%)    & $24478+i0.46$~~
\end{tabular}
\end{ruledtabular}
\end{table}

\section{Epilogue}
\label{sec:summary}

The $S$-wave fully-charm and -bottom pentaquarks with spin-parity $J^P=\frac{1}{2}^-$, $\frac{3}{2}^-$ and $\frac{5}{2}^-$ have been systematically investigated within both real- and complex-scaling range using a Lattice-QCD inspired potential embedded in a quark model picture. The same formalism was applied with success to the fully-charm tetraquarks at the sight of our results and their comparison the last findings of the LHCb collaboration. The 5-body Sch\"odinger-like equation is solved by expanding the wave function solution in terms of Gaussian basis functions whose ranges are in geometric progression. Moreover, the baryon-meson and diquark-diquark-antiquark configurations, along with all of their possible color channels and coupligns, are also comprehensively considered.

Several narrow resonances are obtained in the complete coupled-channel calculations when the complex scaling range allows us to discern them. They are summarized in Table~\ref{GresultCCT}. All these states are exotic color configurations, either hidden-color baryon-meson or diquark-diquark-antiquark structure, or by the coupling between them. As mentioned, their decay widths are generally small but other possible baryon-meson decay channels with higher partial waves have not been considered, neither their three-body decay channels and potentially large electromagnetic transitions. Besides, they are usually compact arrangements of quarks whose sizes are about $0.5-0.7$~fm and $0.2-0.4$~fm for the fully-charm and -bottom pentaquarks, respectively. The compact feature is also indicated in several studies of heavy-flavored multiquark systems; therefore, the predicted fully-heavy pentaquarks are expected to be observed in future experiments, but not without difficulties.


\begin{acknowledgments}
Work partially financed by: the Zhejiang Provincial Natural Science Foundation under Grant No. LQ22A050004; National Natural Science Foundation of China under Grant Nos. 11535005 and 11775118; the Ministerio Espa\~nol de Ciencia e Innovaci\'on under grant No. PID2019-107844GB-C22; and Junta de Andaluc\'ia, contract nos. P18-FR-5057 and Operativo FEDER Andaluc\'ia 2014-2020 UHU-1264517.
\end{acknowledgments}


\bibliography{FHP}

\begin{thebibliography}{107}%
\makeatletter
\providecommand \@ifxundefined [1]{%
 \@ifx{#1\undefined}
}%
\providecommand \@ifnum [1]{%
 \ifnum #1\expandafter \@firstoftwo
 \else \expandafter \@secondoftwo
 \fi
}%
\providecommand \@ifx [1]{%
 \ifx #1\expandafter \@firstoftwo
 \else \expandafter \@secondoftwo
 \fi
}%
\providecommand \natexlab [1]{#1}%
\providecommand \enquote  [1]{``#1''}%
\providecommand \bibnamefont  [1]{#1}%
\providecommand \bibfnamefont [1]{#1}%
\providecommand \citenamefont [1]{#1}%
\providecommand \href@noop [0]{\@secondoftwo}%
\providecommand \href [0]{\begingroup \@sanitize@url \@href}%
\providecommand \@href[1]{\@@startlink{#1}\@@href}%
\providecommand \@@href[1]{\endgroup#1\@@endlink}%
\providecommand \@sanitize@url [0]{\catcode `\\12\catcode `\$12\catcode
  `\&12\catcode `\#12\catcode `\^12\catcode `\_12\catcode `\%12\relax}%
\providecommand \@@startlink[1]{}%
\providecommand \@@endlink[0]{}%
\providecommand \url  [0]{\begingroup\@sanitize@url \@url }%
\providecommand \@url [1]{\endgroup\@href {#1}{\urlprefix }}%
\providecommand \urlprefix  [0]{URL }%
\providecommand \Eprint [0]{\href }%
\providecommand \doibase [0]{http://dx.doi.org/}%
\providecommand \selectlanguage [0]{\@gobble}%
\providecommand \bibinfo  [0]{\@secondoftwo}%
\providecommand \bibfield  [0]{\@secondoftwo}%
\providecommand \translation [1]{[#1]}%
\providecommand \BibitemOpen [0]{}%
\providecommand \bibitemStop [0]{}%
\providecommand \bibitemNoStop [0]{.\EOS\space}%
\providecommand \EOS [0]{\spacefactor3000\relax}%
\providecommand \BibitemShut  [1]{\csname bibitem#1\endcsname}%
\let\auto@bib@innerbib\@empty
\bibitem [{\citenamefont {Gell-Mann}(1964)}]{Gell-Mann:1964ewy}%
  \BibitemOpen
  \bibfield  {author} {\bibinfo {author} {\bibfnamefont {M.}~\bibnamefont
  {Gell-Mann}},\ }\href {\doibase 10.1016/S0031-9163(64)92001-3} {\bibfield
  {journal} {\bibinfo  {journal} {Phys. Lett.}\ }\textbf {\bibinfo {volume}
  {8}},\ \bibinfo {pages} {214} (\bibinfo {year} {1964})}\BibitemShut {NoStop}%
\bibitem [{\citenamefont {Zweig}(1964)}]{Zweig:1964CERN}%
  \BibitemOpen
  \bibfield  {author} {\bibinfo {author} {\bibfnamefont {G.}~\bibnamefont
  {Zweig}},\ }\href@noop {} {\bibfield  {journal} {\bibinfo  {journal} {CERN
  Report No.8182/TH.401, CERN Report No.8419/TH.412}\ } (\bibinfo {year}
  {1964})}\BibitemShut {NoStop}%
\bibitem [{\citenamefont {Choi}\ \emph {et~al.}(2003)\citenamefont {Choi} \emph
  {et~al.}}]{Belle:2003nnu}%
  \BibitemOpen
  \bibfield  {author} {\bibinfo {author} {\bibfnamefont {S.~K.}\ \bibnamefont
  {Choi}} \emph {et~al.} (\bibinfo {collaboration} {Belle}),\ }\href {\doibase
  10.1103/PhysRevLett.91.262001} {\bibfield  {journal} {\bibinfo  {journal}
  {Phys. Rev. Lett.}\ }\textbf {\bibinfo {volume} {91}},\ \bibinfo {pages}
  {262001} (\bibinfo {year} {2003})},\ \Eprint
  {http://arxiv.org/abs/hep-ex/0309032} {arXiv:hep-ex/0309032} \BibitemShut
  {NoStop}%
\bibitem [{\citenamefont {Brambilla}\ \emph {et~al.}(2011)\citenamefont
  {Brambilla} \emph {et~al.}}]{Brambilla:2010cs}%
  \BibitemOpen
  \bibfield  {author} {\bibinfo {author} {\bibfnamefont {N.}~\bibnamefont
  {Brambilla}} \emph {et~al.},\ }\href {\doibase
  10.1140/epjc/s10052-010-1534-9} {\bibfield  {journal} {\bibinfo  {journal}
  {Eur. Phys. J. C}\ }\textbf {\bibinfo {volume} {71}},\ \bibinfo {pages}
  {1534} (\bibinfo {year} {2011})},\ \Eprint {http://arxiv.org/abs/1010.5827}
  {arXiv:1010.5827 [hep-ph]} \BibitemShut {NoStop}%
\bibitem [{\citenamefont {Brambilla}\ \emph {et~al.}(2014)\citenamefont
  {Brambilla} \emph {et~al.}}]{Brambilla:2014jmp}%
  \BibitemOpen
  \bibfield  {author} {\bibinfo {author} {\bibfnamefont {N.}~\bibnamefont
  {Brambilla}} \emph {et~al.},\ }\href {\doibase
  10.1140/epjc/s10052-014-2981-5} {\bibfield  {journal} {\bibinfo  {journal}
  {Eur. Phys. J. C}\ }\textbf {\bibinfo {volume} {74}},\ \bibinfo {pages}
  {2981} (\bibinfo {year} {2014})},\ \Eprint {http://arxiv.org/abs/1404.3723}
  {arXiv:1404.3723 [hep-ph]} \BibitemShut {NoStop}%
\bibitem [{\citenamefont {Olsen}(2015)}]{Olsen:2014qna}%
  \BibitemOpen
  \bibfield  {author} {\bibinfo {author} {\bibfnamefont {S.~L.}\ \bibnamefont
  {Olsen}},\ }\href {\doibase 10.1007/S11467-014-0449-6} {\bibfield  {journal}
  {\bibinfo  {journal} {Front. Phys. (Beijing)}\ }\textbf {\bibinfo {volume}
  {10}},\ \bibinfo {pages} {121} (\bibinfo {year} {2015})},\ \Eprint
  {http://arxiv.org/abs/1411.7738} {arXiv:1411.7738 [hep-ex]} \BibitemShut
  {NoStop}%
\bibitem [{\citenamefont {Esposito}\ \emph {et~al.}(2017)\citenamefont
  {Esposito}, \citenamefont {Pilloni},\ and\ \citenamefont
  {Polosa}}]{Esposito:2016noz}%
  \BibitemOpen
  \bibfield  {author} {\bibinfo {author} {\bibfnamefont {A.}~\bibnamefont
  {Esposito}}, \bibinfo {author} {\bibfnamefont {A.}~\bibnamefont {Pilloni}}, \
  and\ \bibinfo {author} {\bibfnamefont {A.~D.}\ \bibnamefont {Polosa}},\
  }\href {\doibase 10.1016/j.physrep.2016.11.002} {\bibfield  {journal}
  {\bibinfo  {journal} {Phys. Rept.}\ }\textbf {\bibinfo {volume} {668}},\
  \bibinfo {pages} {1} (\bibinfo {year} {2017})},\ \Eprint
  {http://arxiv.org/abs/1611.07920} {arXiv:1611.07920 [hep-ph]} \BibitemShut
  {NoStop}%
\bibitem [{\citenamefont {Chen}\ \emph {et~al.}(2016)\citenamefont {Chen},
  \citenamefont {Chen}, \citenamefont {Liu},\ and\ \citenamefont
  {Zhu}}]{Chen:2016qju}%
  \BibitemOpen
  \bibfield  {author} {\bibinfo {author} {\bibfnamefont {H.-X.}\ \bibnamefont
  {Chen}}, \bibinfo {author} {\bibfnamefont {W.}~\bibnamefont {Chen}}, \bibinfo
  {author} {\bibfnamefont {X.}~\bibnamefont {Liu}}, \ and\ \bibinfo {author}
  {\bibfnamefont {S.-L.}\ \bibnamefont {Zhu}},\ }\href {\doibase
  10.1016/j.physrep.2016.05.004} {\bibfield  {journal} {\bibinfo  {journal}
  {Phys. Rept.}\ }\textbf {\bibinfo {volume} {639}},\ \bibinfo {pages} {1}
  (\bibinfo {year} {2016})},\ \Eprint {http://arxiv.org/abs/1601.02092}
  {arXiv:1601.02092 [hep-ph]} \BibitemShut {NoStop}%
\bibitem [{\citenamefont {Chen}\ \emph
  {et~al.}(2017{\natexlab{a}})\citenamefont {Chen}, \citenamefont {Chen},
  \citenamefont {Liu}, \citenamefont {Liu},\ and\ \citenamefont
  {Zhu}}]{Chen:2016spr}%
  \BibitemOpen
  \bibfield  {author} {\bibinfo {author} {\bibfnamefont {H.-X.}\ \bibnamefont
  {Chen}}, \bibinfo {author} {\bibfnamefont {W.}~\bibnamefont {Chen}}, \bibinfo
  {author} {\bibfnamefont {X.}~\bibnamefont {Liu}}, \bibinfo {author}
  {\bibfnamefont {Y.-R.}\ \bibnamefont {Liu}}, \ and\ \bibinfo {author}
  {\bibfnamefont {S.-L.}\ \bibnamefont {Zhu}},\ }\href {\doibase
  10.1088/1361-6633/aa6420} {\bibfield  {journal} {\bibinfo  {journal} {Rept.
  Prog. Phys.}\ }\textbf {\bibinfo {volume} {80}},\ \bibinfo {pages} {076201}
  (\bibinfo {year} {2017}{\natexlab{a}})},\ \Eprint
  {http://arxiv.org/abs/1609.08928} {arXiv:1609.08928 [hep-ph]} \BibitemShut
  {NoStop}%
\bibitem [{\citenamefont {Karliner}\ \emph {et~al.}(2018)\citenamefont
  {Karliner}, \citenamefont {Rosner},\ and\ \citenamefont
  {Skwarnicki}}]{Karliner:2017qhf}%
  \BibitemOpen
  \bibfield  {author} {\bibinfo {author} {\bibfnamefont {M.}~\bibnamefont
  {Karliner}}, \bibinfo {author} {\bibfnamefont {J.~L.}\ \bibnamefont
  {Rosner}}, \ and\ \bibinfo {author} {\bibfnamefont {T.}~\bibnamefont
  {Skwarnicki}},\ }\href {\doibase 10.1146/annurev-nucl-101917-020902}
  {\bibfield  {journal} {\bibinfo  {journal} {Ann. Rev. Nucl. Part. Sci.}\
  }\textbf {\bibinfo {volume} {68}},\ \bibinfo {pages} {17} (\bibinfo {year}
  {2018})},\ \Eprint {http://arxiv.org/abs/1711.10626} {arXiv:1711.10626
  [hep-ph]} \BibitemShut {NoStop}%
\bibitem [{\citenamefont {Ali}\ \emph {et~al.}(2017)\citenamefont {Ali},
  \citenamefont {Lange},\ and\ \citenamefont {Stone}}]{Ali:2017jda}%
  \BibitemOpen
  \bibfield  {author} {\bibinfo {author} {\bibfnamefont {A.}~\bibnamefont
  {Ali}}, \bibinfo {author} {\bibfnamefont {J.~S.}\ \bibnamefont {Lange}}, \
  and\ \bibinfo {author} {\bibfnamefont {S.}~\bibnamefont {Stone}},\ }\href
  {\doibase 10.1016/j.ppnp.2017.08.003} {\bibfield  {journal} {\bibinfo
  {journal} {Prog. Part. Nucl. Phys.}\ }\textbf {\bibinfo {volume} {97}},\
  \bibinfo {pages} {123} (\bibinfo {year} {2017})},\ \Eprint
  {http://arxiv.org/abs/1706.00610} {arXiv:1706.00610 [hep-ph]} \BibitemShut
  {NoStop}%
\bibitem [{\citenamefont {Guo}\ \emph {et~al.}(2018)\citenamefont {Guo},
  \citenamefont {Hanhart}, \citenamefont {Mei\ss{}ner}, \citenamefont {Wang},
  \citenamefont {Zhao},\ and\ \citenamefont {Zou}}]{Guo:2017jvc}%
  \BibitemOpen
  \bibfield  {author} {\bibinfo {author} {\bibfnamefont {F.-K.}\ \bibnamefont
  {Guo}}, \bibinfo {author} {\bibfnamefont {C.}~\bibnamefont {Hanhart}},
  \bibinfo {author} {\bibfnamefont {U.-G.}\ \bibnamefont {Mei\ss{}ner}},
  \bibinfo {author} {\bibfnamefont {Q.}~\bibnamefont {Wang}}, \bibinfo {author}
  {\bibfnamefont {Q.}~\bibnamefont {Zhao}}, \ and\ \bibinfo {author}
  {\bibfnamefont {B.-S.}\ \bibnamefont {Zou}},\ }\href {\doibase
  10.1103/RevModPhys.90.015004} {\bibfield  {journal} {\bibinfo  {journal}
  {Rev. Mod. Phys.}\ }\textbf {\bibinfo {volume} {90}},\ \bibinfo {pages}
  {015004} (\bibinfo {year} {2018})},\ \Eprint
  {http://arxiv.org/abs/1705.00141} {arXiv:1705.00141 [hep-ph]} \BibitemShut
  {NoStop}%
\bibitem [{\citenamefont {Liu}\ \emph {et~al.}(2019)\citenamefont {Liu},
  \citenamefont {Chen}, \citenamefont {Chen}, \citenamefont {Liu},\ and\
  \citenamefont {Zhu}}]{Liu:2019zoy}%
  \BibitemOpen
  \bibfield  {author} {\bibinfo {author} {\bibfnamefont {Y.-R.}\ \bibnamefont
  {Liu}}, \bibinfo {author} {\bibfnamefont {H.-X.}\ \bibnamefont {Chen}},
  \bibinfo {author} {\bibfnamefont {W.}~\bibnamefont {Chen}}, \bibinfo {author}
  {\bibfnamefont {X.}~\bibnamefont {Liu}}, \ and\ \bibinfo {author}
  {\bibfnamefont {S.-L.}\ \bibnamefont {Zhu}},\ }\href {\doibase
  10.1016/j.ppnp.2019.04.003} {\bibfield  {journal} {\bibinfo  {journal} {Prog.
  Part. Nucl. Phys.}\ }\textbf {\bibinfo {volume} {107}},\ \bibinfo {pages}
  {237} (\bibinfo {year} {2019})},\ \Eprint {http://arxiv.org/abs/1903.11976}
  {arXiv:1903.11976 [hep-ph]} \BibitemShut {NoStop}%
\bibitem [{\citenamefont {Yang}\ \emph
  {et~al.}(2020{\natexlab{a}})\citenamefont {Yang}, \citenamefont {Ping},\ and\
  \citenamefont {Segovia}}]{Yang:2020atz}%
  \BibitemOpen
  \bibfield  {author} {\bibinfo {author} {\bibfnamefont {G.}~\bibnamefont
  {Yang}}, \bibinfo {author} {\bibfnamefont {J.}~\bibnamefont {Ping}}, \ and\
  \bibinfo {author} {\bibfnamefont {J.}~\bibnamefont {Segovia}},\ }\href
  {\doibase 10.3390/sym12111869} {\bibfield  {journal} {\bibinfo  {journal}
  {Symmetry}\ }\textbf {\bibinfo {volume} {12}},\ \bibinfo {pages} {1869}
  (\bibinfo {year} {2020}{\natexlab{a}})},\ \Eprint
  {http://arxiv.org/abs/2009.00238} {arXiv:2009.00238 [hep-ph]} \BibitemShut
  {NoStop}%
\bibitem [{\citenamefont {Dong}\ \emph
  {et~al.}(2021{\natexlab{a}})\citenamefont {Dong}, \citenamefont {Guo},\ and\
  \citenamefont {Zou}}]{Dong:2020hxe}%
  \BibitemOpen
  \bibfield  {author} {\bibinfo {author} {\bibfnamefont {X.-K.}\ \bibnamefont
  {Dong}}, \bibinfo {author} {\bibfnamefont {F.-K.}\ \bibnamefont {Guo}}, \
  and\ \bibinfo {author} {\bibfnamefont {B.-S.}\ \bibnamefont {Zou}},\ }\href
  {\doibase 10.1103/PhysRevLett.126.152001} {\bibfield  {journal} {\bibinfo
  {journal} {Phys. Rev. Lett.}\ }\textbf {\bibinfo {volume} {126}},\ \bibinfo
  {pages} {152001} (\bibinfo {year} {2021}{\natexlab{a}})},\ \Eprint
  {http://arxiv.org/abs/2011.14517} {arXiv:2011.14517 [hep-ph]} \BibitemShut
  {NoStop}%
\bibitem [{\citenamefont {Dong}\ \emph
  {et~al.}(2021{\natexlab{b}})\citenamefont {Dong}, \citenamefont {Guo},\ and\
  \citenamefont {Zou}}]{Dong:2021bvy}%
  \BibitemOpen
  \bibfield  {author} {\bibinfo {author} {\bibfnamefont {X.-K.}\ \bibnamefont
  {Dong}}, \bibinfo {author} {\bibfnamefont {F.-K.}\ \bibnamefont {Guo}}, \
  and\ \bibinfo {author} {\bibfnamefont {B.-S.}\ \bibnamefont {Zou}},\ }\href
  {\doibase 10.1088/1572-9494/ac27a2} {\bibfield  {journal} {\bibinfo
  {journal} {Commun. Theor. Phys.}\ }\textbf {\bibinfo {volume} {73}},\
  \bibinfo {pages} {125201} (\bibinfo {year} {2021}{\natexlab{b}})},\ \Eprint
  {http://arxiv.org/abs/2108.02673} {arXiv:2108.02673 [hep-ph]} \BibitemShut
  {NoStop}%
\bibitem [{\citenamefont {Chen}(2022)}]{Chen:2021erj}%
  \BibitemOpen
  \bibfield  {author} {\bibinfo {author} {\bibfnamefont {H.-X.}\ \bibnamefont
  {Chen}},\ }\href {\doibase 10.1103/PhysRevD.105.094003} {\bibfield  {journal}
  {\bibinfo  {journal} {Phys. Rev. D}\ }\textbf {\bibinfo {volume} {105}},\
  \bibinfo {pages} {094003} (\bibinfo {year} {2022})},\ \Eprint
  {http://arxiv.org/abs/2103.08586} {arXiv:2103.08586 [hep-ph]} \BibitemShut
  {NoStop}%
\bibitem [{\citenamefont {Meng}\ \emph {et~al.}(2022)\citenamefont {Meng},
  \citenamefont {Wang}, \citenamefont {Wang},\ and\ \citenamefont
  {Zhu}}]{Meng:2022ozq}%
  \BibitemOpen
  \bibfield  {author} {\bibinfo {author} {\bibfnamefont {L.}~\bibnamefont
  {Meng}}, \bibinfo {author} {\bibfnamefont {B.}~\bibnamefont {Wang}}, \bibinfo
  {author} {\bibfnamefont {G.-J.}\ \bibnamefont {Wang}}, \ and\ \bibinfo
  {author} {\bibfnamefont {S.-L.}\ \bibnamefont {Zhu}},\ }\href@noop {} {\
  (\bibinfo {year} {2022})},\ \Eprint {http://arxiv.org/abs/2204.08716}
  {arXiv:2204.08716 [hep-ph]} \BibitemShut {NoStop}%
\bibitem [{\citenamefont {Aaij}\ \emph
  {et~al.}(2020{\natexlab{a}})\citenamefont {Aaij} \emph
  {et~al.}}]{LHCb:2020pxc}%
  \BibitemOpen
  \bibfield  {author} {\bibinfo {author} {\bibfnamefont {R.}~\bibnamefont
  {Aaij}} \emph {et~al.} (\bibinfo {collaboration} {LHCb}),\ }\href {\doibase
  10.1103/PhysRevD.102.112003} {\bibfield  {journal} {\bibinfo  {journal}
  {Phys. Rev. D}\ }\textbf {\bibinfo {volume} {102}},\ \bibinfo {pages}
  {112003} (\bibinfo {year} {2020}{\natexlab{a}})},\ \Eprint
  {http://arxiv.org/abs/2009.00026} {arXiv:2009.00026 [hep-ex]} \BibitemShut
  {NoStop}%
\bibitem [{\citenamefont {Aaij}\ \emph
  {et~al.}(2020{\natexlab{b}})\citenamefont {Aaij} \emph
  {et~al.}}]{LHCb:2020bls}%
  \BibitemOpen
  \bibfield  {author} {\bibinfo {author} {\bibfnamefont {R.}~\bibnamefont
  {Aaij}} \emph {et~al.} (\bibinfo {collaboration} {LHCb}),\ }\href {\doibase
  10.1103/PhysRevLett.125.242001} {\bibfield  {journal} {\bibinfo  {journal}
  {Phys. Rev. Lett.}\ }\textbf {\bibinfo {volume} {125}},\ \bibinfo {pages}
  {242001} (\bibinfo {year} {2020}{\natexlab{b}})},\ \Eprint
  {http://arxiv.org/abs/2009.00025} {arXiv:2009.00025 [hep-ex]} \BibitemShut
  {NoStop}%
\bibitem [{\citenamefont {Ablikim}\ \emph
  {et~al.}(2022{\natexlab{a}})\citenamefont {Ablikim} \emph
  {et~al.}}]{BESIII:2022llk}%
  \BibitemOpen
  \bibfield  {author} {\bibinfo {author} {\bibfnamefont {M.}~\bibnamefont
  {Ablikim}} \emph {et~al.} (\bibinfo {collaboration} {BESIII}),\ }\href@noop
  {} {\  (\bibinfo {year} {2022}{\natexlab{a}})},\ \Eprint
  {http://arxiv.org/abs/2201.10796} {arXiv:2201.10796 [hep-ex]} \BibitemShut
  {NoStop}%
\bibitem [{\citenamefont {Ablikim}\ \emph
  {et~al.}(2021{\natexlab{a}})\citenamefont {Ablikim} \emph
  {et~al.}}]{BESIII:2021xoh}%
  \BibitemOpen
  \bibfield  {author} {\bibinfo {author} {\bibfnamefont {M.}~\bibnamefont
  {Ablikim}} \emph {et~al.} (\bibinfo {collaboration} {BESIII}),\ }\href@noop
  {} {\  (\bibinfo {year} {2021}{\natexlab{a}})},\ \Eprint
  {http://arxiv.org/abs/2112.14369} {arXiv:2112.14369 [hep-ex]} \BibitemShut
  {NoStop}%
\bibitem [{\citenamefont {Ablikim}\ \emph
  {et~al.}(2021{\natexlab{b}})\citenamefont {Ablikim} \emph
  {et~al.}}]{BESIII:2020qkh}%
  \BibitemOpen
  \bibfield  {author} {\bibinfo {author} {\bibfnamefont {M.}~\bibnamefont
  {Ablikim}} \emph {et~al.} (\bibinfo {collaboration} {BESIII}),\ }\href
  {\doibase 10.1103/PhysRevLett.126.102001} {\bibfield  {journal} {\bibinfo
  {journal} {Phys. Rev. Lett.}\ }\textbf {\bibinfo {volume} {126}},\ \bibinfo
  {pages} {102001} (\bibinfo {year} {2021}{\natexlab{b}})},\ \Eprint
  {http://arxiv.org/abs/2011.07855} {arXiv:2011.07855 [hep-ex]} \BibitemShut
  {NoStop}%
\bibitem [{\citenamefont {Aaij}\ \emph
  {et~al.}(2021{\natexlab{a}})\citenamefont {Aaij} \emph
  {et~al.}}]{LHCb:2021uow}%
  \BibitemOpen
  \bibfield  {author} {\bibinfo {author} {\bibfnamefont {R.}~\bibnamefont
  {Aaij}} \emph {et~al.} (\bibinfo {collaboration} {LHCb}),\ }\href {\doibase
  10.1103/PhysRevLett.127.082001} {\bibfield  {journal} {\bibinfo  {journal}
  {Phys. Rev. Lett.}\ }\textbf {\bibinfo {volume} {127}},\ \bibinfo {pages}
  {082001} (\bibinfo {year} {2021}{\natexlab{a}})},\ \Eprint
  {http://arxiv.org/abs/2103.01803} {arXiv:2103.01803 [hep-ex]} \BibitemShut
  {NoStop}%
\bibitem [{\citenamefont {Ablikim}\ \emph
  {et~al.}(2022{\natexlab{b}})\citenamefont {Ablikim} \emph
  {et~al.}}]{BESIII:2022joj}%
  \BibitemOpen
  \bibfield  {author} {\bibinfo {author} {\bibfnamefont {M.}~\bibnamefont
  {Ablikim}} \emph {et~al.} (\bibinfo {collaboration} {BESIII}),\ }\href@noop
  {} {\  (\bibinfo {year} {2022}{\natexlab{b}})},\ \Eprint
  {http://arxiv.org/abs/2204.07800} {arXiv:2204.07800 [hep-ex]} \BibitemShut
  {NoStop}%
\bibitem [{\citenamefont {Aaij}\ \emph
  {et~al.}(2022{\natexlab{a}})\citenamefont {Aaij} \emph
  {et~al.}}]{LHCb:2022kpe}%
  \BibitemOpen
  \bibfield  {author} {\bibinfo {author} {\bibfnamefont {R.}~\bibnamefont
  {Aaij}} \emph {et~al.} (\bibinfo {collaboration} {LHCb}),\ }\href@noop {} {\
  (\bibinfo {year} {2022}{\natexlab{a}})},\ \Eprint
  {http://arxiv.org/abs/2202.04045} {arXiv:2202.04045 [hep-ex]} \BibitemShut
  {NoStop}%
\bibitem [{\citenamefont {Ablikim}\ \emph
  {et~al.}(2022{\natexlab{c}})\citenamefont {Ablikim} \emph
  {et~al.}}]{BESIII:2022yga}%
  \BibitemOpen
  \bibfield  {author} {\bibinfo {author} {\bibfnamefont {M.}~\bibnamefont
  {Ablikim}} \emph {et~al.} (\bibinfo {collaboration} {BESIII}),\ }\href@noop
  {} {\  (\bibinfo {year} {2022}{\natexlab{c}})},\ \Eprint
  {http://arxiv.org/abs/2203.05815} {arXiv:2203.05815 [hep-ex]} \BibitemShut
  {NoStop}%
\bibitem [{\citenamefont {Aaij}\ \emph
  {et~al.}(2021{\natexlab{b}})\citenamefont {Aaij} \emph
  {et~al.}}]{LHCb:2021vvq}%
  \BibitemOpen
  \bibfield  {author} {\bibinfo {author} {\bibfnamefont {R.}~\bibnamefont
  {Aaij}} \emph {et~al.} (\bibinfo {collaboration} {LHCb}),\ }\href@noop {} {\
  (\bibinfo {year} {2021}{\natexlab{b}})},\ \Eprint
  {http://arxiv.org/abs/2109.01038} {arXiv:2109.01038 [hep-ex]} \BibitemShut
  {NoStop}%
\bibitem [{\citenamefont {Aaij}\ \emph
  {et~al.}(2021{\natexlab{c}})\citenamefont {Aaij} \emph
  {et~al.}}]{LHCb:2021auc}%
  \BibitemOpen
  \bibfield  {author} {\bibinfo {author} {\bibfnamefont {R.}~\bibnamefont
  {Aaij}} \emph {et~al.} (\bibinfo {collaboration} {LHCb}),\ }\href@noop {} {\
  (\bibinfo {year} {2021}{\natexlab{c}})},\ \Eprint
  {http://arxiv.org/abs/2109.01056} {arXiv:2109.01056 [hep-ex]} \BibitemShut
  {NoStop}%
\bibitem [{\citenamefont {Aaij}\ \emph {et~al.}(2015)\citenamefont {Aaij} \emph
  {et~al.}}]{Aaij:2015tga}%
  \BibitemOpen
  \bibfield  {author} {\bibinfo {author} {\bibfnamefont {R.}~\bibnamefont
  {Aaij}} \emph {et~al.} (\bibinfo {collaboration} {LHCb}),\ }\href {\doibase
  10.1103/PhysRevLett.115.072001} {\bibfield  {journal} {\bibinfo  {journal}
  {Phys. Rev. Lett.}\ }\textbf {\bibinfo {volume} {115}},\ \bibinfo {pages}
  {072001} (\bibinfo {year} {2015})}\BibitemShut {NoStop}%
\bibitem [{\citenamefont {Aaij}\ \emph {et~al.}(2019)\citenamefont {Aaij} \emph
  {et~al.}}]{lhcb:2019pc}%
  \BibitemOpen
  \bibfield  {author} {\bibinfo {author} {\bibfnamefont {R.}~\bibnamefont
  {Aaij}} \emph {et~al.} (\bibinfo {collaboration} {LHCb}),\ }\href {\doibase
  10.1103/PhysRevLett.122.222001} {\bibfield  {journal} {\bibinfo  {journal}
  {Phys. Rev. Lett.}\ }\textbf {\bibinfo {volume} {122}},\ \bibinfo {pages}
  {222001} (\bibinfo {year} {2019})}\BibitemShut {NoStop}%
\bibitem [{\citenamefont {Aaij}\ \emph
  {et~al.}(2022{\natexlab{b}})\citenamefont {Aaij} \emph
  {et~al.}}]{LHCb:2021chn}%
  \BibitemOpen
  \bibfield  {author} {\bibinfo {author} {\bibfnamefont {R.}~\bibnamefont
  {Aaij}} \emph {et~al.} (\bibinfo {collaboration} {LHCb}),\ }\href {\doibase
  10.1103/PhysRevLett.128.062001} {\bibfield  {journal} {\bibinfo  {journal}
  {Phys. Rev. Lett.}\ }\textbf {\bibinfo {volume} {128}},\ \bibinfo {pages}
  {062001} (\bibinfo {year} {2022}{\natexlab{b}})},\ \Eprint
  {http://arxiv.org/abs/2108.04720} {arXiv:2108.04720 [hep-ex]} \BibitemShut
  {NoStop}%
\bibitem [{\citenamefont {Aaij}\ \emph
  {et~al.}(2021{\natexlab{d}})\citenamefont {Aaij} \emph
  {et~al.}}]{LHCb:2020jpq}%
  \BibitemOpen
  \bibfield  {author} {\bibinfo {author} {\bibfnamefont {R.}~\bibnamefont
  {Aaij}} \emph {et~al.} (\bibinfo {collaboration} {LHCb}),\ }\href {\doibase
  10.1016/j.scib.2021.02.030} {\bibfield  {journal} {\bibinfo  {journal} {Sci.
  Bull.}\ }\textbf {\bibinfo {volume} {66}},\ \bibinfo {pages} {1391} (\bibinfo
  {year} {2021}{\natexlab{d}})},\ \Eprint {http://arxiv.org/abs/2012.10380}
  {arXiv:2012.10380 [hep-ex]} \BibitemShut {NoStop}%
\bibitem [{\citenamefont {Weinberg}(2013)}]{Weinberg:2013cfa}%
  \BibitemOpen
  \bibfield  {author} {\bibinfo {author} {\bibfnamefont {S.}~\bibnamefont
  {Weinberg}},\ }\href {\doibase 10.1103/PhysRevLett.110.261601} {\bibfield
  {journal} {\bibinfo  {journal} {Phys. Rev. Lett.}\ }\textbf {\bibinfo
  {volume} {110}},\ \bibinfo {pages} {261601} (\bibinfo {year} {2013})},\
  \Eprint {http://arxiv.org/abs/1303.0342} {arXiv:1303.0342 [hep-ph]}
  \BibitemShut {NoStop}%
\bibitem [{\citenamefont {Braaten}\ \emph {et~al.}(2014)\citenamefont
  {Braaten}, \citenamefont {Langmack},\ and\ \citenamefont
  {Smith}}]{Braaten:2014qka}%
  \BibitemOpen
  \bibfield  {author} {\bibinfo {author} {\bibfnamefont {E.}~\bibnamefont
  {Braaten}}, \bibinfo {author} {\bibfnamefont {C.}~\bibnamefont {Langmack}}, \
  and\ \bibinfo {author} {\bibfnamefont {D.~H.}\ \bibnamefont {Smith}},\ }\href
  {\doibase 10.1103/PhysRevD.90.014044} {\bibfield  {journal} {\bibinfo
  {journal} {Phys. Rev. D}\ }\textbf {\bibinfo {volume} {90}},\ \bibinfo
  {pages} {014044} (\bibinfo {year} {2014})},\ \Eprint
  {http://arxiv.org/abs/1402.0438} {arXiv:1402.0438 [hep-ph]} \BibitemShut
  {NoStop}%
\bibitem [{\citenamefont {Brodsky}\ and\ \citenamefont
  {Lebed}(2015)}]{Brodsky:2015wza}%
  \BibitemOpen
  \bibfield  {author} {\bibinfo {author} {\bibfnamefont {S.~J.}\ \bibnamefont
  {Brodsky}}\ and\ \bibinfo {author} {\bibfnamefont {R.~F.}\ \bibnamefont
  {Lebed}},\ }\href {\doibase 10.1103/PhysRevD.91.114025} {\bibfield  {journal}
  {\bibinfo  {journal} {Phys. Rev. D}\ }\textbf {\bibinfo {volume} {91}},\
  \bibinfo {pages} {114025} (\bibinfo {year} {2015})},\ \Eprint
  {http://arxiv.org/abs/1505.00803} {arXiv:1505.00803 [hep-ph]} \BibitemShut
  {NoStop}%
\bibitem [{\citenamefont {Chen}\ \emph
  {et~al.}(2017{\natexlab{b}})\citenamefont {Chen}, \citenamefont {Chen},
  \citenamefont {Liu}, \citenamefont {Steele},\ and\ \citenamefont
  {Zhu}}]{Chen:2016jxd}%
  \BibitemOpen
  \bibfield  {author} {\bibinfo {author} {\bibfnamefont {W.}~\bibnamefont
  {Chen}}, \bibinfo {author} {\bibfnamefont {H.-X.}\ \bibnamefont {Chen}},
  \bibinfo {author} {\bibfnamefont {X.}~\bibnamefont {Liu}}, \bibinfo {author}
  {\bibfnamefont {T.~G.}\ \bibnamefont {Steele}}, \ and\ \bibinfo {author}
  {\bibfnamefont {S.-L.}\ \bibnamefont {Zhu}},\ }\href {\doibase
  10.1016/j.physletb.2017.08.034} {\bibfield  {journal} {\bibinfo  {journal}
  {Phys. Lett. B}\ }\textbf {\bibinfo {volume} {773}},\ \bibinfo {pages} {247}
  (\bibinfo {year} {2017}{\natexlab{b}})},\ \Eprint
  {http://arxiv.org/abs/1605.01647} {arXiv:1605.01647 [hep-ph]} \BibitemShut
  {NoStop}%
\bibitem [{\citenamefont {Eichten}\ and\ \citenamefont
  {Quigg}(2017)}]{Eichten:2017ffp}%
  \BibitemOpen
  \bibfield  {author} {\bibinfo {author} {\bibfnamefont {E.~J.}\ \bibnamefont
  {Eichten}}\ and\ \bibinfo {author} {\bibfnamefont {C.}~\bibnamefont
  {Quigg}},\ }\href {\doibase 10.1103/PhysRevLett.119.202002} {\bibfield
  {journal} {\bibinfo  {journal} {Phys. Rev. Lett.}\ }\textbf {\bibinfo
  {volume} {119}},\ \bibinfo {pages} {202002} (\bibinfo {year} {2017})},\
  \Eprint {http://arxiv.org/abs/1707.09575} {arXiv:1707.09575 [hep-ph]}
  \BibitemShut {NoStop}%
\bibitem [{\citenamefont {Richard}\ \emph {et~al.}(2018)\citenamefont
  {Richard}, \citenamefont {Valcarce},\ and\ \citenamefont
  {Vijande}}]{Richard:2018yrm}%
  \BibitemOpen
  \bibfield  {author} {\bibinfo {author} {\bibfnamefont {J.-M.}\ \bibnamefont
  {Richard}}, \bibinfo {author} {\bibfnamefont {A.}~\bibnamefont {Valcarce}}, \
  and\ \bibinfo {author} {\bibfnamefont {J.}~\bibnamefont {Vijande}},\ }\href
  {\doibase 10.1103/PhysRevC.97.035211} {\bibfield  {journal} {\bibinfo
  {journal} {Phys. Rev. C}\ }\textbf {\bibinfo {volume} {97}},\ \bibinfo
  {pages} {035211} (\bibinfo {year} {2018})},\ \Eprint
  {http://arxiv.org/abs/1803.06155} {arXiv:1803.06155 [hep-ph]} \BibitemShut
  {NoStop}%
\bibitem [{\citenamefont {Ortega}\ \emph {et~al.}(2019)\citenamefont {Ortega},
  \citenamefont {Segovia}, \citenamefont {Entem},\ and\ \citenamefont
  {Fern\'andez}}]{Ortega:2018cnm}%
  \BibitemOpen
  \bibfield  {author} {\bibinfo {author} {\bibfnamefont {P.~G.}\ \bibnamefont
  {Ortega}}, \bibinfo {author} {\bibfnamefont {J.}~\bibnamefont {Segovia}},
  \bibinfo {author} {\bibfnamefont {D.~R.}\ \bibnamefont {Entem}}, \ and\
  \bibinfo {author} {\bibfnamefont {F.}~\bibnamefont {Fern\'andez}},\ }\href
  {\doibase 10.1140/epjc/s10052-019-6552-7} {\bibfield  {journal} {\bibinfo
  {journal} {Eur. Phys. J. C}\ }\textbf {\bibinfo {volume} {79}},\ \bibinfo
  {pages} {78} (\bibinfo {year} {2019})},\ \Eprint
  {http://arxiv.org/abs/1808.00914} {arXiv:1808.00914 [hep-ph]} \BibitemShut
  {NoStop}%
\bibitem [{\citenamefont {Bedolla}\ \emph {et~al.}(2020)\citenamefont
  {Bedolla}, \citenamefont {Ferretti}, \citenamefont {Roberts},\ and\
  \citenamefont {Santopinto}}]{Bedolla:2019zwg}%
  \BibitemOpen
  \bibfield  {author} {\bibinfo {author} {\bibfnamefont {M.~A.}\ \bibnamefont
  {Bedolla}}, \bibinfo {author} {\bibfnamefont {J.}~\bibnamefont {Ferretti}},
  \bibinfo {author} {\bibfnamefont {C.~D.}\ \bibnamefont {Roberts}}, \ and\
  \bibinfo {author} {\bibfnamefont {E.}~\bibnamefont {Santopinto}},\ }\href
  {\doibase 10.1140/epjc/s10052-020-08579-3} {\bibfield  {journal} {\bibinfo
  {journal} {Eur. Phys. J. C}\ }\textbf {\bibinfo {volume} {80}},\ \bibinfo
  {pages} {1004} (\bibinfo {year} {2020})},\ \Eprint
  {http://arxiv.org/abs/1911.00960} {arXiv:1911.00960 [hep-ph]} \BibitemShut
  {NoStop}%
\bibitem [{\citenamefont {Ortega}\ \emph {et~al.}(2021)\citenamefont {Ortega},
  \citenamefont {Segovia},\ and\ \citenamefont {Fernandez}}]{Ortega:2021xst}%
  \BibitemOpen
  \bibfield  {author} {\bibinfo {author} {\bibfnamefont {P.~G.}\ \bibnamefont
  {Ortega}}, \bibinfo {author} {\bibfnamefont {J.}~\bibnamefont {Segovia}}, \
  and\ \bibinfo {author} {\bibfnamefont {F.}~\bibnamefont {Fernandez}},\ }\href
  {\doibase 10.1103/PhysRevD.104.094004} {\bibfield  {journal} {\bibinfo
  {journal} {Phys. Rev. D}\ }\textbf {\bibinfo {volume} {104}},\ \bibinfo
  {pages} {094004} (\bibinfo {year} {2021})},\ \Eprint
  {http://arxiv.org/abs/2107.02544} {arXiv:2107.02544 [hep-ph]} \BibitemShut
  {NoStop}%
\bibitem [{\citenamefont {Esposito}\ \emph {et~al.}(2021)\citenamefont
  {Esposito}, \citenamefont {Manzari}, \citenamefont {Pilloni},\ and\
  \citenamefont {Polosa}}]{Esposito:2021ptx}%
  \BibitemOpen
  \bibfield  {author} {\bibinfo {author} {\bibfnamefont {A.}~\bibnamefont
  {Esposito}}, \bibinfo {author} {\bibfnamefont {C.~A.}\ \bibnamefont
  {Manzari}}, \bibinfo {author} {\bibfnamefont {A.}~\bibnamefont {Pilloni}}, \
  and\ \bibinfo {author} {\bibfnamefont {A.~D.}\ \bibnamefont {Polosa}},\
  }\href {\doibase 10.1103/PhysRevD.104.114029} {\bibfield  {journal} {\bibinfo
   {journal} {Phys. Rev. D}\ }\textbf {\bibinfo {volume} {104}},\ \bibinfo
  {pages} {114029} (\bibinfo {year} {2021})},\ \Eprint
  {http://arxiv.org/abs/2109.10359} {arXiv:2109.10359 [hep-ph]} \BibitemShut
  {NoStop}%
\bibitem [{\citenamefont {Balassa}\ and\ \citenamefont
  {Wolf}(2021)}]{Balassa:2021ssx}%
  \BibitemOpen
  \bibfield  {author} {\bibinfo {author} {\bibfnamefont {G.}~\bibnamefont
  {Balassa}}\ and\ \bibinfo {author} {\bibfnamefont {G.}~\bibnamefont {Wolf}},\
  }\href {\doibase 10.1140/epja/s10050-021-00553-1} {\bibfield  {journal}
  {\bibinfo  {journal} {Eur. Phys. J. A}\ }\textbf {\bibinfo {volume} {57}},\
  \bibinfo {pages} {246} (\bibinfo {year} {2021})},\ \Eprint
  {http://arxiv.org/abs/2111.05602} {arXiv:2111.05602 [hep-ph]} \BibitemShut
  {NoStop}%
\bibitem [{\citenamefont {Cheng}\ \emph {et~al.}(2021)\citenamefont {Cheng},
  \citenamefont {Yang},\ and\ \citenamefont {Huang}}]{Cheng:2021gca}%
  \BibitemOpen
  \bibfield  {author} {\bibinfo {author} {\bibfnamefont {C.}~\bibnamefont
  {Cheng}}, \bibinfo {author} {\bibfnamefont {F.}~\bibnamefont {Yang}}, \ and\
  \bibinfo {author} {\bibfnamefont {Y.}~\bibnamefont {Huang}},\ }\href
  {\doibase 10.1103/PhysRevD.104.116007} {\bibfield  {journal} {\bibinfo
  {journal} {Phys. Rev. D}\ }\textbf {\bibinfo {volume} {104}},\ \bibinfo
  {pages} {116007} (\bibinfo {year} {2021})},\ \Eprint
  {http://arxiv.org/abs/2110.04746} {arXiv:2110.04746 [hep-ph]} \BibitemShut
  {NoStop}%
\bibitem [{\citenamefont {Du}\ \emph {et~al.}(2021{\natexlab{a}})\citenamefont
  {Du}, \citenamefont {Guo},\ and\ \citenamefont {Oller}}]{Du:2021bgb}%
  \BibitemOpen
  \bibfield  {author} {\bibinfo {author} {\bibfnamefont {M.-L.}\ \bibnamefont
  {Du}}, \bibinfo {author} {\bibfnamefont {Z.-H.}\ \bibnamefont {Guo}}, \ and\
  \bibinfo {author} {\bibfnamefont {J.~A.}\ \bibnamefont {Oller}},\ }\href
  {\doibase 10.1103/PhysRevD.104.114034} {\bibfield  {journal} {\bibinfo
  {journal} {Phys. Rev. D}\ }\textbf {\bibinfo {volume} {104}},\ \bibinfo
  {pages} {114034} (\bibinfo {year} {2021}{\natexlab{a}})},\ \Eprint
  {http://arxiv.org/abs/2109.14237} {arXiv:2109.14237 [hep-ph]} \BibitemShut
  {NoStop}%
\bibitem [{\citenamefont {Yalikun}\ \emph {et~al.}(2021)\citenamefont
  {Yalikun}, \citenamefont {Lin}, \citenamefont {Guo}, \citenamefont {Kamiya},\
  and\ \citenamefont {Zou}}]{Yalikun:2021bfm}%
  \BibitemOpen
  \bibfield  {author} {\bibinfo {author} {\bibfnamefont {N.}~\bibnamefont
  {Yalikun}}, \bibinfo {author} {\bibfnamefont {Y.-H.}\ \bibnamefont {Lin}},
  \bibinfo {author} {\bibfnamefont {F.-K.}\ \bibnamefont {Guo}}, \bibinfo
  {author} {\bibfnamefont {Y.}~\bibnamefont {Kamiya}}, \ and\ \bibinfo {author}
  {\bibfnamefont {B.-S.}\ \bibnamefont {Zou}},\ }\href {\doibase
  10.1103/PhysRevD.104.094039} {\bibfield  {journal} {\bibinfo  {journal}
  {Phys. Rev. D}\ }\textbf {\bibinfo {volume} {104}},\ \bibinfo {pages}
  {094039} (\bibinfo {year} {2021})},\ \Eprint
  {http://arxiv.org/abs/2109.03504} {arXiv:2109.03504 [hep-ph]} \BibitemShut
  {NoStop}%
\bibitem [{\citenamefont {Wang}\ \emph {et~al.}(2021)\citenamefont {Wang},
  \citenamefont {Liu},\ and\ \citenamefont {Matsuki}}]{Wang:2021crr}%
  \BibitemOpen
  \bibfield  {author} {\bibinfo {author} {\bibfnamefont {J.-Z.}\ \bibnamefont
  {Wang}}, \bibinfo {author} {\bibfnamefont {X.}~\bibnamefont {Liu}}, \ and\
  \bibinfo {author} {\bibfnamefont {T.}~\bibnamefont {Matsuki}},\ }\href
  {\doibase 10.1103/PhysRevD.104.114020} {\bibfield  {journal} {\bibinfo
  {journal} {Phys. Rev. D}\ }\textbf {\bibinfo {volume} {104}},\ \bibinfo
  {pages} {114020} (\bibinfo {year} {2021})},\ \Eprint
  {http://arxiv.org/abs/2110.09423} {arXiv:2110.09423 [hep-ph]} \BibitemShut
  {NoStop}%
\bibitem [{\citenamefont {Chen}\ \emph
  {et~al.}(2021{\natexlab{a}})\citenamefont {Chen}, \citenamefont {Xie},
  \citenamefont {Xu}, \citenamefont {Zhang}, \citenamefont {Zhou},
  \citenamefont {She},\ and\ \citenamefont {Chen}}]{Chen:2021obo}%
  \BibitemOpen
  \bibfield  {author} {\bibinfo {author} {\bibfnamefont {C.-h.}\ \bibnamefont
  {Chen}}, \bibinfo {author} {\bibfnamefont {Y.-L.}\ \bibnamefont {Xie}},
  \bibinfo {author} {\bibfnamefont {H.-g.}\ \bibnamefont {Xu}}, \bibinfo
  {author} {\bibfnamefont {Z.}~\bibnamefont {Zhang}}, \bibinfo {author}
  {\bibfnamefont {D.-M.}\ \bibnamefont {Zhou}}, \bibinfo {author}
  {\bibfnamefont {Z.-L.}\ \bibnamefont {She}}, \ and\ \bibinfo {author}
  {\bibfnamefont {G.}~\bibnamefont {Chen}},\ }\href@noop {} {\  (\bibinfo
  {year} {2021}{\natexlab{a}})},\ \Eprint {http://arxiv.org/abs/2111.03241}
  {arXiv:2111.03241 [hep-ph]} \BibitemShut {NoStop}%
\bibitem [{\citenamefont {Du}(2022)}]{Du:2021fmt}%
  \BibitemOpen
  \bibfield  {author} {\bibinfo {author} {\bibfnamefont {M.-L.}\ \bibnamefont
  {Du}},\ }\href {\doibase 10.1051/epjconf/202225804007} {\bibfield  {journal}
  {\bibinfo  {journal} {EPJ Web Conf.}\ }\textbf {\bibinfo {volume} {258}},\
  \bibinfo {pages} {04007} (\bibinfo {year} {2022})},\ \Eprint
  {http://arxiv.org/abs/2111.11405} {arXiv:2111.11405 [hep-ph]} \BibitemShut
  {NoStop}%
\bibitem [{\citenamefont {Xie}\ \emph {et~al.}(2022)\citenamefont {Xie},
  \citenamefont {Ling}, \citenamefont {Liu},\ and\ \citenamefont
  {Geng}}]{Xie:2022hhv}%
  \BibitemOpen
  \bibfield  {author} {\bibinfo {author} {\bibfnamefont {J.-M.}\ \bibnamefont
  {Xie}}, \bibinfo {author} {\bibfnamefont {X.-Z.}\ \bibnamefont {Ling}},
  \bibinfo {author} {\bibfnamefont {M.-Z.}\ \bibnamefont {Liu}}, \ and\
  \bibinfo {author} {\bibfnamefont {L.-S.}\ \bibnamefont {Geng}},\ }\href@noop
  {} {\  (\bibinfo {year} {2022})},\ \Eprint {http://arxiv.org/abs/2204.12356}
  {arXiv:2204.12356 [hep-ph]} \BibitemShut {NoStop}%
\bibitem [{\citenamefont {Wang}\ \emph {et~al.}(2022)\citenamefont {Wang},
  \citenamefont {Shen}, \citenamefont {R\"onchen}, \citenamefont
  {Mei\ss{}ner},\ and\ \citenamefont {Zou}}]{Wang:2022oof}%
  \BibitemOpen
  \bibfield  {author} {\bibinfo {author} {\bibfnamefont {Z.-L.}\ \bibnamefont
  {Wang}}, \bibinfo {author} {\bibfnamefont {C.-W.}\ \bibnamefont {Shen}},
  \bibinfo {author} {\bibfnamefont {D.}~\bibnamefont {R\"onchen}}, \bibinfo
  {author} {\bibfnamefont {U.-G.}\ \bibnamefont {Mei\ss{}ner}}, \ and\ \bibinfo
  {author} {\bibfnamefont {B.-S.}\ \bibnamefont {Zou}},\ }\href@noop {} {\
  (\bibinfo {year} {2022})},\ \Eprint {http://arxiv.org/abs/2204.12122}
  {arXiv:2204.12122 [hep-ph]} \BibitemShut {NoStop}%
\bibitem [{\citenamefont {Deng}(2022)}]{Deng:2022vkv}%
  \BibitemOpen
  \bibfield  {author} {\bibinfo {author} {\bibfnamefont {C.-R.}\ \bibnamefont
  {Deng}},\ }\href@noop {} {\  (\bibinfo {year} {2022})},\ \Eprint
  {http://arxiv.org/abs/2202.13570} {arXiv:2202.13570 [hep-ph]} \BibitemShut
  {NoStop}%
\bibitem [{\citenamefont {Park}\ \emph {et~al.}(2022)\citenamefont {Park},
  \citenamefont {Lee}, \citenamefont {Cho},\ and\ \citenamefont
  {Kim}}]{Park:2022nza}%
  \BibitemOpen
  \bibfield  {author} {\bibinfo {author} {\bibfnamefont {I.~W.}\ \bibnamefont
  {Park}}, \bibinfo {author} {\bibfnamefont {S.~H.}\ \bibnamefont {Lee}},
  \bibinfo {author} {\bibfnamefont {S.}~\bibnamefont {Cho}}, \ and\ \bibinfo
  {author} {\bibfnamefont {Y.}~\bibnamefont {Kim}},\ }\href@noop {} {\
  (\bibinfo {year} {2022})},\ \Eprint {http://arxiv.org/abs/2202.11631}
  {arXiv:2202.11631 [hep-ph]} \BibitemShut {NoStop}%
\bibitem [{\citenamefont {Chen}\ and\ \citenamefont
  {Liu}(2022)}]{Chen:2022onm}%
  \BibitemOpen
  \bibfield  {author} {\bibinfo {author} {\bibfnamefont {R.}~\bibnamefont
  {Chen}}\ and\ \bibinfo {author} {\bibfnamefont {X.}~\bibnamefont {Liu}},\
  }\href {\doibase 10.1103/PhysRevD.105.014029} {\bibfield  {journal} {\bibinfo
   {journal} {Phys. Rev. D}\ }\textbf {\bibinfo {volume} {105}},\ \bibinfo
  {pages} {014029} (\bibinfo {year} {2022})},\ \Eprint
  {http://arxiv.org/abs/2201.07603} {arXiv:2201.07603 [hep-ph]} \BibitemShut
  {NoStop}%
\bibitem [{\citenamefont {Burns}\ and\ \citenamefont
  {Swanson}(2022)}]{Burns:2021jlu}%
  \BibitemOpen
  \bibfield  {author} {\bibinfo {author} {\bibfnamefont {T.~J.}\ \bibnamefont
  {Burns}}\ and\ \bibinfo {author} {\bibfnamefont {E.~S.}\ \bibnamefont
  {Swanson}},\ }\href {\doibase 10.1140/epja/s10050-022-00723-9} {\bibfield
  {journal} {\bibinfo  {journal} {Eur. Phys. J. A}\ }\textbf {\bibinfo {volume}
  {58}},\ \bibinfo {pages} {68} (\bibinfo {year} {2022})},\ \Eprint
  {http://arxiv.org/abs/2112.11527} {arXiv:2112.11527 [hep-ph]} \BibitemShut
  {NoStop}%
\bibitem [{\citenamefont {Yang}\ \emph
  {et~al.}(2021{\natexlab{a}})\citenamefont {Yang}, \citenamefont {Huang},\
  and\ \citenamefont {Zhu}}]{Yang:2021pio}%
  \BibitemOpen
  \bibfield  {author} {\bibinfo {author} {\bibfnamefont {F.}~\bibnamefont
  {Yang}}, \bibinfo {author} {\bibfnamefont {Y.}~\bibnamefont {Huang}}, \ and\
  \bibinfo {author} {\bibfnamefont {H.~Q.}\ \bibnamefont {Zhu}},\ }\href
  {\doibase 10.1007/s11433-021-1796-0} {\bibfield  {journal} {\bibinfo
  {journal} {Sci. China Phys. Mech. Astron.}\ }\textbf {\bibinfo {volume}
  {64}},\ \bibinfo {pages} {121011} (\bibinfo {year} {2021}{\natexlab{a}})},\
  \Eprint {http://arxiv.org/abs/2107.13267} {arXiv:2107.13267 [hep-ph]}
  \BibitemShut {NoStop}%
\bibitem [{\citenamefont {Shi}\ \emph {et~al.}(2021)\citenamefont {Shi},
  \citenamefont {Huang},\ and\ \citenamefont {Wang}}]{Shi:2021wyt}%
  \BibitemOpen
  \bibfield  {author} {\bibinfo {author} {\bibfnamefont {P.-P.}\ \bibnamefont
  {Shi}}, \bibinfo {author} {\bibfnamefont {F.}~\bibnamefont {Huang}}, \ and\
  \bibinfo {author} {\bibfnamefont {W.-L.}\ \bibnamefont {Wang}},\ }\href
  {\doibase 10.1140/epja/s10050-021-00542-4} {\bibfield  {journal} {\bibinfo
  {journal} {Eur. Phys. J. A}\ }\textbf {\bibinfo {volume} {57}},\ \bibinfo
  {pages} {237} (\bibinfo {year} {2021})},\ \Eprint
  {http://arxiv.org/abs/2107.08680} {arXiv:2107.08680 [hep-ph]} \BibitemShut
  {NoStop}%
\bibitem [{\citenamefont {Li}\ \emph {et~al.}(2021)\citenamefont {Li},
  \citenamefont {Liu}, \citenamefont {Sun},\ and\ \citenamefont
  {Chen}}]{Li:2021ryu}%
  \BibitemOpen
  \bibfield  {author} {\bibinfo {author} {\bibfnamefont {M.-W.}\ \bibnamefont
  {Li}}, \bibinfo {author} {\bibfnamefont {Z.-W.}\ \bibnamefont {Liu}},
  \bibinfo {author} {\bibfnamefont {Z.-F.}\ \bibnamefont {Sun}}, \ and\
  \bibinfo {author} {\bibfnamefont {R.}~\bibnamefont {Chen}},\ }\href {\doibase
  10.1103/PhysRevD.104.054016} {\bibfield  {journal} {\bibinfo  {journal}
  {Phys. Rev. D}\ }\textbf {\bibinfo {volume} {104}},\ \bibinfo {pages}
  {054016} (\bibinfo {year} {2021})},\ \Eprint
  {http://arxiv.org/abs/2106.15053} {arXiv:2106.15053 [hep-ph]} \BibitemShut
  {NoStop}%
\bibitem [{\citenamefont {Ling}\ \emph
  {et~al.}(2021{\natexlab{a}})\citenamefont {Ling}, \citenamefont {Lu},
  \citenamefont {Liu},\ and\ \citenamefont {Geng}}]{Ling:2021lmq}%
  \BibitemOpen
  \bibfield  {author} {\bibinfo {author} {\bibfnamefont {X.-Z.}\ \bibnamefont
  {Ling}}, \bibinfo {author} {\bibfnamefont {J.-X.}\ \bibnamefont {Lu}},
  \bibinfo {author} {\bibfnamefont {M.-Z.}\ \bibnamefont {Liu}}, \ and\
  \bibinfo {author} {\bibfnamefont {L.-S.}\ \bibnamefont {Geng}},\ }\href
  {\doibase 10.1103/PhysRevD.104.074022} {\bibfield  {journal} {\bibinfo
  {journal} {Phys. Rev. D}\ }\textbf {\bibinfo {volume} {104}},\ \bibinfo
  {pages} {074022} (\bibinfo {year} {2021}{\natexlab{a}})},\ \Eprint
  {http://arxiv.org/abs/2106.12250} {arXiv:2106.12250 [hep-ph]} \BibitemShut
  {NoStop}%
\bibitem [{\citenamefont {Wu}\ \emph {et~al.}(2021{\natexlab{a}})\citenamefont
  {Wu}, \citenamefont {Pan}, \citenamefont {Liu}, \citenamefont {Lu},
  \citenamefont {Geng},\ and\ \citenamefont {Liu}}]{Wu:2021gyn}%
  \BibitemOpen
  \bibfield  {author} {\bibinfo {author} {\bibfnamefont {T.-W.}\ \bibnamefont
  {Wu}}, \bibinfo {author} {\bibfnamefont {Y.-W.}\ \bibnamefont {Pan}},
  \bibinfo {author} {\bibfnamefont {M.-Z.}\ \bibnamefont {Liu}}, \bibinfo
  {author} {\bibfnamefont {J.-X.}\ \bibnamefont {Lu}}, \bibinfo {author}
  {\bibfnamefont {L.-S.}\ \bibnamefont {Geng}}, \ and\ \bibinfo {author}
  {\bibfnamefont {X.-H.}\ \bibnamefont {Liu}},\ }\href {\doibase
  10.1103/PhysRevD.104.094032} {\bibfield  {journal} {\bibinfo  {journal}
  {Phys. Rev. D}\ }\textbf {\bibinfo {volume} {104}},\ \bibinfo {pages}
  {094032} (\bibinfo {year} {2021}{\natexlab{a}})},\ \Eprint
  {http://arxiv.org/abs/2106.11450} {arXiv:2106.11450 [hep-ph]} \BibitemShut
  {NoStop}%
\bibitem [{\citenamefont {Ruangyoo}\ \emph {et~al.}(2021)\citenamefont
  {Ruangyoo}, \citenamefont {Phumphan}, \citenamefont {Chen}, \citenamefont
  {Limphirat},\ and\ \citenamefont {Yan}}]{Ruangyoo:2021aoi}%
  \BibitemOpen
  \bibfield  {author} {\bibinfo {author} {\bibfnamefont {W.}~\bibnamefont
  {Ruangyoo}}, \bibinfo {author} {\bibfnamefont {K.}~\bibnamefont {Phumphan}},
  \bibinfo {author} {\bibfnamefont {C.-C.}\ \bibnamefont {Chen}}, \bibinfo
  {author} {\bibfnamefont {A.}~\bibnamefont {Limphirat}}, \ and\ \bibinfo
  {author} {\bibfnamefont {Y.}~\bibnamefont {Yan}},\ }\href@noop {} {\
  (\bibinfo {year} {2021})},\ \Eprint {http://arxiv.org/abs/2105.14249}
  {arXiv:2105.14249 [hep-ph]} \BibitemShut {NoStop}%
\bibitem [{\citenamefont {Ling}\ \emph
  {et~al.}(2021{\natexlab{b}})\citenamefont {Ling}, \citenamefont {Dai},
  \citenamefont {Du},\ and\ \citenamefont {Wang}}]{Ling:2021sld}%
  \BibitemOpen
  \bibfield  {author} {\bibinfo {author} {\bibfnamefont {P.}~\bibnamefont
  {Ling}}, \bibinfo {author} {\bibfnamefont {X.-H.}\ \bibnamefont {Dai}},
  \bibinfo {author} {\bibfnamefont {M.-L.}\ \bibnamefont {Du}}, \ and\ \bibinfo
  {author} {\bibfnamefont {Q.}~\bibnamefont {Wang}},\ }\href {\doibase
  10.1140/epjc/s10052-021-09613-8} {\bibfield  {journal} {\bibinfo  {journal}
  {Eur. Phys. J. C}\ }\textbf {\bibinfo {volume} {81}},\ \bibinfo {pages} {819}
  (\bibinfo {year} {2021}{\natexlab{b}})},\ \Eprint
  {http://arxiv.org/abs/2104.11133} {arXiv:2104.11133 [hep-ph]} \BibitemShut
  {NoStop}%
\bibitem [{\citenamefont {Lu}\ \emph {et~al.}(2021)\citenamefont {Lu},
  \citenamefont {Liu}, \citenamefont {Shi},\ and\ \citenamefont
  {Geng}}]{Lu:2021irg}%
  \BibitemOpen
  \bibfield  {author} {\bibinfo {author} {\bibfnamefont {J.-X.}\ \bibnamefont
  {Lu}}, \bibinfo {author} {\bibfnamefont {M.-Z.}\ \bibnamefont {Liu}},
  \bibinfo {author} {\bibfnamefont {R.-X.}\ \bibnamefont {Shi}}, \ and\
  \bibinfo {author} {\bibfnamefont {L.-S.}\ \bibnamefont {Geng}},\ }\href
  {\doibase 10.1103/PhysRevD.104.034022} {\bibfield  {journal} {\bibinfo
  {journal} {Phys. Rev. D}\ }\textbf {\bibinfo {volume} {104}},\ \bibinfo
  {pages} {034022} (\bibinfo {year} {2021})},\ \Eprint
  {http://arxiv.org/abs/2104.10303} {arXiv:2104.10303 [hep-ph]} \BibitemShut
  {NoStop}%
\bibitem [{\citenamefont {Wu}\ \emph {et~al.}(2021{\natexlab{b}})\citenamefont
  {Wu}, \citenamefont {Chen},\ and\ \citenamefont {Ji}}]{Wu:2021caw}%
  \BibitemOpen
  \bibfield  {author} {\bibinfo {author} {\bibfnamefont {Q.}~\bibnamefont
  {Wu}}, \bibinfo {author} {\bibfnamefont {D.-Y.}\ \bibnamefont {Chen}}, \ and\
  \bibinfo {author} {\bibfnamefont {R.}~\bibnamefont {Ji}},\ }\href {\doibase
  10.1088/0256-307X/38/7/071301} {\bibfield  {journal} {\bibinfo  {journal}
  {Chin. Phys. Lett.}\ }\textbf {\bibinfo {volume} {38}},\ \bibinfo {pages}
  {071301} (\bibinfo {year} {2021}{\natexlab{b}})},\ \Eprint
  {http://arxiv.org/abs/2103.05257} {arXiv:2103.05257 [hep-ph]} \BibitemShut
  {NoStop}%
\bibitem [{\citenamefont {Du}\ \emph {et~al.}(2021{\natexlab{b}})\citenamefont
  {Du}, \citenamefont {Baru}, \citenamefont {Guo}, \citenamefont {Hanhart},
  \citenamefont {Mei\ss{}ner}, \citenamefont {Oller},\ and\ \citenamefont
  {Wang}}]{Du:2021fmf}%
  \BibitemOpen
  \bibfield  {author} {\bibinfo {author} {\bibfnamefont {M.-L.}\ \bibnamefont
  {Du}}, \bibinfo {author} {\bibfnamefont {V.}~\bibnamefont {Baru}}, \bibinfo
  {author} {\bibfnamefont {F.-K.}\ \bibnamefont {Guo}}, \bibinfo {author}
  {\bibfnamefont {C.}~\bibnamefont {Hanhart}}, \bibinfo {author} {\bibfnamefont
  {U.-G.}\ \bibnamefont {Mei\ss{}ner}}, \bibinfo {author} {\bibfnamefont
  {J.~A.}\ \bibnamefont {Oller}}, \ and\ \bibinfo {author} {\bibfnamefont
  {Q.}~\bibnamefont {Wang}},\ }\href {\doibase 10.1007/JHEP08(2021)157}
  {\bibfield  {journal} {\bibinfo  {journal} {JHEP}\ }\textbf {\bibinfo
  {volume} {08}},\ \bibinfo {pages} {157} (\bibinfo {year}
  {2021}{\natexlab{b}})},\ \Eprint {http://arxiv.org/abs/2102.07159}
  {arXiv:2102.07159 [hep-ph]} \BibitemShut {NoStop}%
\bibitem [{\citenamefont {Xiao}\ \emph {et~al.}(2021)\citenamefont {Xiao},
  \citenamefont {Wu},\ and\ \citenamefont {Zou}}]{Xiao:2021rgp}%
  \BibitemOpen
  \bibfield  {author} {\bibinfo {author} {\bibfnamefont {C.~W.}\ \bibnamefont
  {Xiao}}, \bibinfo {author} {\bibfnamefont {J.~J.}\ \bibnamefont {Wu}}, \ and\
  \bibinfo {author} {\bibfnamefont {B.~S.}\ \bibnamefont {Zou}},\ }\href
  {\doibase 10.1103/PhysRevD.103.054016} {\bibfield  {journal} {\bibinfo
  {journal} {Phys. Rev. D}\ }\textbf {\bibinfo {volume} {103}},\ \bibinfo
  {pages} {054016} (\bibinfo {year} {2021})},\ \Eprint
  {http://arxiv.org/abs/2102.02607} {arXiv:2102.02607 [hep-ph]} \BibitemShut
  {NoStop}%
\bibitem [{\citenamefont {Zhu}\ \emph {et~al.}(2021)\citenamefont {Zhu},
  \citenamefont {Song},\ and\ \citenamefont {He}}]{Zhu:2021lhd}%
  \BibitemOpen
  \bibfield  {author} {\bibinfo {author} {\bibfnamefont {J.-T.}\ \bibnamefont
  {Zhu}}, \bibinfo {author} {\bibfnamefont {L.-Q.}\ \bibnamefont {Song}}, \
  and\ \bibinfo {author} {\bibfnamefont {J.}~\bibnamefont {He}},\ }\href
  {\doibase 10.1103/PhysRevD.103.074007} {\bibfield  {journal} {\bibinfo
  {journal} {Phys. Rev. D}\ }\textbf {\bibinfo {volume} {103}},\ \bibinfo
  {pages} {074007} (\bibinfo {year} {2021})},\ \Eprint
  {http://arxiv.org/abs/2101.12441} {arXiv:2101.12441 [hep-ph]} \BibitemShut
  {NoStop}%
\bibitem [{\citenamefont {Chen}(2021)}]{Chen:2021tip}%
  \BibitemOpen
  \bibfield  {author} {\bibinfo {author} {\bibfnamefont {R.}~\bibnamefont
  {Chen}},\ }\href {\doibase 10.1140/epjc/s10052-021-08904-4} {\bibfield
  {journal} {\bibinfo  {journal} {Eur. Phys. J. C}\ }\textbf {\bibinfo {volume}
  {81}},\ \bibinfo {pages} {122} (\bibinfo {year} {2021})},\ \Eprint
  {http://arxiv.org/abs/2101.10614} {arXiv:2101.10614 [hep-ph]} \BibitemShut
  {NoStop}%
\bibitem [{\citenamefont {Yan}\ \emph {et~al.}(2021)\citenamefont {Yan},
  \citenamefont {Peng}, \citenamefont {S\'anchez},\ and\ \citenamefont
  {Valderrama}}]{Yan:2021nio}%
  \BibitemOpen
  \bibfield  {author} {\bibinfo {author} {\bibfnamefont {M.-J.}\ \bibnamefont
  {Yan}}, \bibinfo {author} {\bibfnamefont {F.-Z.}\ \bibnamefont {Peng}},
  \bibinfo {author} {\bibfnamefont {M.~S.}\ \bibnamefont {S\'anchez}}, \ and\
  \bibinfo {author} {\bibfnamefont {M.~P.}\ \bibnamefont {Valderrama}},\
  }\href@noop {} {\  (\bibinfo {year} {2021})},\ \Eprint
  {http://arxiv.org/abs/2108.05306} {arXiv:2108.05306 [hep-ph]} \BibitemShut
  {NoStop}%
\bibitem [{\citenamefont {Yang}\ \emph {et~al.}(2017)\citenamefont {Yang},
  \citenamefont {Ping},\ and\ \citenamefont {Wang}}]{Yang:2015bmv}%
  \BibitemOpen
  \bibfield  {author} {\bibinfo {author} {\bibfnamefont {G.}~\bibnamefont
  {Yang}}, \bibinfo {author} {\bibfnamefont {J.}~\bibnamefont {Ping}}, \ and\
  \bibinfo {author} {\bibfnamefont {F.}~\bibnamefont {Wang}},\ }\href {\doibase
  10.1103/PhysRevD.95.014010} {\bibfield  {journal} {\bibinfo  {journal} {Phys.
  Rev.}\ }\textbf {\bibinfo {volume} {D95}},\ \bibinfo {pages} {014010}
  (\bibinfo {year} {2017})}\BibitemShut {NoStop}%
\bibitem [{\citenamefont {Yang}\ and\ \citenamefont
  {Chen}(2022)}]{Yang:2022uot}%
  \BibitemOpen
  \bibfield  {author} {\bibinfo {author} {\bibfnamefont {P.}~\bibnamefont
  {Yang}}\ and\ \bibinfo {author} {\bibfnamefont {W.}~\bibnamefont {Chen}},\
  }\href@noop {} {\  (\bibinfo {year} {2022})},\ \Eprint
  {http://arxiv.org/abs/2203.15616} {arXiv:2203.15616 [hep-ph]} \BibitemShut
  {NoStop}%
\bibitem [{\citenamefont {Huang}\ and\ \citenamefont
  {Zhu}(2021)}]{Huang:2021ave}%
  \BibitemOpen
  \bibfield  {author} {\bibinfo {author} {\bibfnamefont {Y.}~\bibnamefont
  {Huang}}\ and\ \bibinfo {author} {\bibfnamefont {H.~Q.}\ \bibnamefont
  {Zhu}},\ }\href {\doibase 10.1103/PhysRevD.104.056027} {\bibfield  {journal}
  {\bibinfo  {journal} {Phys. Rev. D}\ }\textbf {\bibinfo {volume} {104}},\
  \bibinfo {pages} {056027} (\bibinfo {year} {2021})},\ \Eprint
  {http://arxiv.org/abs/2107.03773} {arXiv:2107.03773 [hep-ph]} \BibitemShut
  {NoStop}%
\bibitem [{\citenamefont {Zhu}\ \emph {et~al.}(2020)\citenamefont {Zhu},
  \citenamefont {Kong}, \citenamefont {Liu},\ and\ \citenamefont
  {He}}]{Zhu:2020vto}%
  \BibitemOpen
  \bibfield  {author} {\bibinfo {author} {\bibfnamefont {J.-T.}\ \bibnamefont
  {Zhu}}, \bibinfo {author} {\bibfnamefont {S.-Y.}\ \bibnamefont {Kong}},
  \bibinfo {author} {\bibfnamefont {Y.}~\bibnamefont {Liu}}, \ and\ \bibinfo
  {author} {\bibfnamefont {J.}~\bibnamefont {He}},\ }\href {\doibase
  10.1140/epjc/s10052-020-8410-z} {\bibfield  {journal} {\bibinfo  {journal}
  {Eur. Phys. J. C}\ }\textbf {\bibinfo {volume} {80}},\ \bibinfo {pages}
  {1016} (\bibinfo {year} {2020})},\ \Eprint {http://arxiv.org/abs/2007.07596}
  {arXiv:2007.07596 [hep-ph]} \BibitemShut {NoStop}%
\bibitem [{\citenamefont {Yang}\ \emph {et~al.}(2019)\citenamefont {Yang},
  \citenamefont {Ping},\ and\ \citenamefont {Segovia}}]{Yang:2018oqd}%
  \BibitemOpen
  \bibfield  {author} {\bibinfo {author} {\bibfnamefont {G.}~\bibnamefont
  {Yang}}, \bibinfo {author} {\bibfnamefont {J.}~\bibnamefont {Ping}}, \ and\
  \bibinfo {author} {\bibfnamefont {J.}~\bibnamefont {Segovia}},\ }\href
  {\doibase 10.1103/PhysRevD.99.014035} {\bibfield  {journal} {\bibinfo
  {journal} {Phys. Rev.}\ }\textbf {\bibinfo {volume} {D99}},\ \bibinfo {pages}
  {014035} (\bibinfo {year} {2019})}\BibitemShut {NoStop}%
\bibitem [{\citenamefont {Zhang}\ \emph {et~al.}(2021)\citenamefont {Zhang},
  \citenamefont {Hu}, \citenamefont {He},\ and\ \citenamefont
  {Ping}}]{Zhang:2020dwp}%
  \BibitemOpen
  \bibfield  {author} {\bibinfo {author} {\bibfnamefont {Q.}~\bibnamefont
  {Zhang}}, \bibinfo {author} {\bibfnamefont {X.-H.}\ \bibnamefont {Hu}},
  \bibinfo {author} {\bibfnamefont {B.-R.}\ \bibnamefont {He}}, \ and\ \bibinfo
  {author} {\bibfnamefont {J.-L.}\ \bibnamefont {Ping}},\ }\href {\doibase
  10.1140/epjc/s10052-021-09017-8} {\bibfield  {journal} {\bibinfo  {journal}
  {Eur. Phys. J. C}\ }\textbf {\bibinfo {volume} {81}},\ \bibinfo {pages} {224}
  (\bibinfo {year} {2021})},\ \Eprint {http://arxiv.org/abs/2012.02017}
  {arXiv:2012.02017 [hep-ph]} \BibitemShut {NoStop}%
\bibitem [{\citenamefont {Xing}\ \emph {et~al.}(2022)\citenamefont {Xing},
  \citenamefont {Liu},\ and\ \citenamefont {Xiao}}]{Xing:2022aij}%
  \BibitemOpen
  \bibfield  {author} {\bibinfo {author} {\bibfnamefont {Y.}~\bibnamefont
  {Xing}}, \bibinfo {author} {\bibfnamefont {W.-L.}\ \bibnamefont {Liu}}, \
  and\ \bibinfo {author} {\bibfnamefont {Y.-H.}\ \bibnamefont {Xiao}},\
  }\href@noop {} {\  (\bibinfo {year} {2022})},\ \Eprint
  {http://arxiv.org/abs/2203.03248} {arXiv:2203.03248 [hep-ph]} \BibitemShut
  {NoStop}%
\bibitem [{\citenamefont {\"Ozdem}(2022)}]{Ozdem:2022vip}%
  \BibitemOpen
  \bibfield  {author} {\bibinfo {author} {\bibfnamefont {U.}~\bibnamefont
  {\"Ozdem}},\ }\href@noop {} {\  (\bibinfo {year} {2022})},\ \Eprint
  {http://arxiv.org/abs/2201.00979} {arXiv:2201.00979 [hep-ph]} \BibitemShut
  {NoStop}%
\bibitem [{\citenamefont {Xing}\ and\ \citenamefont
  {Niu}(2021)}]{Xing:2021yid}%
  \BibitemOpen
  \bibfield  {author} {\bibinfo {author} {\bibfnamefont {Y.}~\bibnamefont
  {Xing}}\ and\ \bibinfo {author} {\bibfnamefont {Y.}~\bibnamefont {Niu}},\
  }\href {\doibase 10.1140/epjc/s10052-021-09730-4} {\bibfield  {journal}
  {\bibinfo  {journal} {Eur. Phys. J. C}\ }\textbf {\bibinfo {volume} {81}},\
  \bibinfo {pages} {978} (\bibinfo {year} {2021})},\ \Eprint
  {http://arxiv.org/abs/2106.09939} {arXiv:2106.09939 [hep-ph]} \BibitemShut
  {NoStop}%
\bibitem [{\citenamefont {Chen}\ \emph
  {et~al.}(2021{\natexlab{b}})\citenamefont {Chen}, \citenamefont {Li},
  \citenamefont {Sun}, \citenamefont {Liu},\ and\ \citenamefont
  {Zhu}}]{Chen:2021kad}%
  \BibitemOpen
  \bibfield  {author} {\bibinfo {author} {\bibfnamefont {R.}~\bibnamefont
  {Chen}}, \bibinfo {author} {\bibfnamefont {N.}~\bibnamefont {Li}}, \bibinfo
  {author} {\bibfnamefont {Z.-F.}\ \bibnamefont {Sun}}, \bibinfo {author}
  {\bibfnamefont {X.}~\bibnamefont {Liu}}, \ and\ \bibinfo {author}
  {\bibfnamefont {S.-L.}\ \bibnamefont {Zhu}},\ }\href {\doibase
  10.1016/j.physletb.2021.136693} {\bibfield  {journal} {\bibinfo  {journal}
  {Phys. Lett. B}\ }\textbf {\bibinfo {volume} {822}},\ \bibinfo {pages}
  {136693} (\bibinfo {year} {2021}{\natexlab{b}})},\ \Eprint
  {http://arxiv.org/abs/2108.12730} {arXiv:2108.12730 [hep-ph]} \BibitemShut
  {NoStop}%
\bibitem [{\citenamefont {Chen}\ \emph
  {et~al.}(2021{\natexlab{c}})\citenamefont {Chen}, \citenamefont {Wang},\ and\
  \citenamefont {Zhu}}]{Chen:2021htr}%
  \BibitemOpen
  \bibfield  {author} {\bibinfo {author} {\bibfnamefont {K.}~\bibnamefont
  {Chen}}, \bibinfo {author} {\bibfnamefont {B.}~\bibnamefont {Wang}}, \ and\
  \bibinfo {author} {\bibfnamefont {S.-L.}\ \bibnamefont {Zhu}},\ }\href
  {\doibase 10.1103/PhysRevD.103.116017} {\bibfield  {journal} {\bibinfo
  {journal} {Phys. Rev. D}\ }\textbf {\bibinfo {volume} {103}},\ \bibinfo
  {pages} {116017} (\bibinfo {year} {2021}{\natexlab{c}})},\ \Eprint
  {http://arxiv.org/abs/2102.05868} {arXiv:2102.05868 [hep-ph]} \BibitemShut
  {NoStop}%
\bibitem [{\citenamefont {Yang}\ \emph
  {et~al.}(2020{\natexlab{b}})\citenamefont {Yang}, \citenamefont {Ping},\ and\
  \citenamefont {Segovia}}]{gy:2020dcp}%
  \BibitemOpen
  \bibfield  {author} {\bibinfo {author} {\bibfnamefont {G.}~\bibnamefont
  {Yang}}, \bibinfo {author} {\bibfnamefont {J.~L.}\ \bibnamefont {Ping}}, \
  and\ \bibinfo {author} {\bibfnamefont {J.}~\bibnamefont {Segovia}},\ }\href
  {\doibase 10.1103/PhysRevD.101.074030} {\bibfield  {journal} {\bibinfo
  {journal} {Phys. Rev. D}\ }\textbf {\bibinfo {volume} {101}},\ \bibinfo
  {pages} {074030} (\bibinfo {year} {2020}{\natexlab{b}})}\BibitemShut
  {NoStop}%
\bibitem [{\citenamefont {Khachatryan}\ \emph {et~al.}(2017)\citenamefont
  {Khachatryan} \emph {et~al.}}]{CMS:2016liw}%
  \BibitemOpen
  \bibfield  {author} {\bibinfo {author} {\bibfnamefont {V.}~\bibnamefont
  {Khachatryan}} \emph {et~al.} (\bibinfo {collaboration} {CMS}),\ }\href
  {\doibase 10.1007/JHEP05(2017)013} {\bibfield  {journal} {\bibinfo  {journal}
  {JHEP}\ }\textbf {\bibinfo {volume} {05}},\ \bibinfo {pages} {013} (\bibinfo
  {year} {2017})},\ \Eprint {http://arxiv.org/abs/1610.07095} {arXiv:1610.07095
  [hep-ex]} \BibitemShut {NoStop}%
\bibitem [{\citenamefont {Yi}(2019)}]{Yi:2018fxo}%
  \BibitemOpen
  \bibfield  {author} {\bibinfo {author} {\bibfnamefont {K.}~\bibnamefont
  {Yi}},\ }\href {\doibase 10.1142/S0217751X1850224X} {\bibfield  {journal}
  {\bibinfo  {journal} {Int. J. Mod. Phys. A}\ }\textbf {\bibinfo {volume}
  {33}},\ \bibinfo {pages} {1850224} (\bibinfo {year} {2019})},\ \Eprint
  {http://arxiv.org/abs/1806.08398} {arXiv:1806.08398 [hep-ph]} \BibitemShut
  {NoStop}%
\bibitem [{\citenamefont {Bland}\ \emph {et~al.}(2019)\citenamefont {Bland}
  \emph {et~al.}}]{ANDY:2019bfn}%
  \BibitemOpen
  \bibfield  {author} {\bibinfo {author} {\bibfnamefont {L.~C.}\ \bibnamefont
  {Bland}} \emph {et~al.} (\bibinfo {collaboration} {ANDY}),\ }\href@noop {} {\
   (\bibinfo {year} {2019})},\ \Eprint {http://arxiv.org/abs/1909.03124}
  {arXiv:1909.03124 [nucl-ex]} \BibitemShut {NoStop}%
\bibitem [{\citenamefont {Aaij}\ \emph {et~al.}(2018)\citenamefont {Aaij} \emph
  {et~al.}}]{LHCb:2018uwm}%
  \BibitemOpen
  \bibfield  {author} {\bibinfo {author} {\bibfnamefont {R.}~\bibnamefont
  {Aaij}} \emph {et~al.} (\bibinfo {collaboration} {LHCb}),\ }\href {\doibase
  10.1007/JHEP10(2018)086} {\bibfield  {journal} {\bibinfo  {journal} {JHEP}\
  }\textbf {\bibinfo {volume} {10}},\ \bibinfo {pages} {086} (\bibinfo {year}
  {2018})},\ \Eprint {http://arxiv.org/abs/1806.09707} {arXiv:1806.09707
  [hep-ex]} \BibitemShut {NoStop}%
\bibitem [{\citenamefont {Sirunyan}\ \emph {et~al.}(2020)\citenamefont
  {Sirunyan} \emph {et~al.}}]{CMS:2020qwa}%
  \BibitemOpen
  \bibfield  {author} {\bibinfo {author} {\bibfnamefont {A.~M.}\ \bibnamefont
  {Sirunyan}} \emph {et~al.} (\bibinfo {collaboration} {CMS}),\ }\href
  {\doibase 10.1016/j.physletb.2020.135578} {\bibfield  {journal} {\bibinfo
  {journal} {Phys. Lett. B}\ }\textbf {\bibinfo {volume} {808}},\ \bibinfo
  {pages} {135578} (\bibinfo {year} {2020})},\ \Eprint
  {http://arxiv.org/abs/2002.06393} {arXiv:2002.06393 [hep-ex]} \BibitemShut
  {NoStop}%
\bibitem [{\citenamefont {Aaij}\ \emph
  {et~al.}(2020{\natexlab{c}})\citenamefont {Aaij} \emph
  {et~al.}}]{LHCb:2020bwg}%
  \BibitemOpen
  \bibfield  {author} {\bibinfo {author} {\bibfnamefont {R.}~\bibnamefont
  {Aaij}} \emph {et~al.} (\bibinfo {collaboration} {LHCb}),\ }\href {\doibase
  10.1016/j.scib.2020.08.032} {\bibfield  {journal} {\bibinfo  {journal} {Sci.
  Bull.}\ }\textbf {\bibinfo {volume} {65}},\ \bibinfo {pages} {1983} (\bibinfo
  {year} {2020}{\natexlab{c}})},\ \Eprint {http://arxiv.org/abs/2006.16957}
  {arXiv:2006.16957 [hep-ex]} \BibitemShut {NoStop}%
\bibitem [{\citenamefont {Yan}\ \emph {et~al.}(2022)\citenamefont {Yan},
  \citenamefont {Wu}, \citenamefont {Hu}, \citenamefont {Huang},\ and\
  \citenamefont {Ping}}]{Yan:2021glh}%
  \BibitemOpen
  \bibfield  {author} {\bibinfo {author} {\bibfnamefont {Y.}~\bibnamefont
  {Yan}}, \bibinfo {author} {\bibfnamefont {Y.}~\bibnamefont {Wu}}, \bibinfo
  {author} {\bibfnamefont {X.}~\bibnamefont {Hu}}, \bibinfo {author}
  {\bibfnamefont {H.}~\bibnamefont {Huang}}, \ and\ \bibinfo {author}
  {\bibfnamefont {J.}~\bibnamefont {Ping}},\ }\href {\doibase
  10.1103/PhysRevD.105.014027} {\bibfield  {journal} {\bibinfo  {journal}
  {Phys. Rev. D}\ }\textbf {\bibinfo {volume} {105}},\ \bibinfo {pages}
  {014027} (\bibinfo {year} {2022})},\ \Eprint
  {http://arxiv.org/abs/2110.10853} {arXiv:2110.10853 [hep-ph]} \BibitemShut
  {NoStop}%
\bibitem [{\citenamefont {An}\ \emph {et~al.}(2021)\citenamefont {An},
  \citenamefont {Chen}, \citenamefont {Liu},\ and\ \citenamefont
  {Liu}}]{An:2020jix}%
  \BibitemOpen
  \bibfield  {author} {\bibinfo {author} {\bibfnamefont {H.-T.}\ \bibnamefont
  {An}}, \bibinfo {author} {\bibfnamefont {K.}~\bibnamefont {Chen}}, \bibinfo
  {author} {\bibfnamefont {Z.-W.}\ \bibnamefont {Liu}}, \ and\ \bibinfo
  {author} {\bibfnamefont {X.}~\bibnamefont {Liu}},\ }\href {\doibase
  10.1103/PhysRevD.103.074006} {\bibfield  {journal} {\bibinfo  {journal}
  {Phys. Rev. D}\ }\textbf {\bibinfo {volume} {103}},\ \bibinfo {pages}
  {074006} (\bibinfo {year} {2021})},\ \Eprint
  {http://arxiv.org/abs/2012.12459} {arXiv:2012.12459 [hep-ph]} \BibitemShut
  {NoStop}%
\bibitem [{\citenamefont {An}\ \emph {et~al.}(2022)\citenamefont {An},
  \citenamefont {Luo}, \citenamefont {Liu},\ and\ \citenamefont
  {Liu}}]{An:2022fvs}%
  \BibitemOpen
  \bibfield  {author} {\bibinfo {author} {\bibfnamefont {H.-T.}\ \bibnamefont
  {An}}, \bibinfo {author} {\bibfnamefont {S.-Q.}\ \bibnamefont {Luo}},
  \bibinfo {author} {\bibfnamefont {Z.-W.}\ \bibnamefont {Liu}}, \ and\
  \bibinfo {author} {\bibfnamefont {X.}~\bibnamefont {Liu}},\ }\href@noop {} {\
   (\bibinfo {year} {2022})},\ \Eprint {http://arxiv.org/abs/2203.03448}
  {arXiv:2203.03448 [hep-ph]} \BibitemShut {NoStop}%
\bibitem [{\citenamefont {Wang}(2021)}]{Wang:2021xao}%
  \BibitemOpen
  \bibfield  {author} {\bibinfo {author} {\bibfnamefont {Z.-G.}\ \bibnamefont
  {Wang}},\ }\href {\doibase 10.1016/j.nuclphysb.2021.115579} {\bibfield
  {journal} {\bibinfo  {journal} {Nucl. Phys. B}\ }\textbf {\bibinfo {volume}
  {973}},\ \bibinfo {pages} {115579} (\bibinfo {year} {2021})},\ \Eprint
  {http://arxiv.org/abs/2104.12090} {arXiv:2104.12090 [hep-ph]} \BibitemShut
  {NoStop}%
\bibitem [{\citenamefont {Zhang}(2021)}]{Zhang:2020vpz}%
  \BibitemOpen
  \bibfield  {author} {\bibinfo {author} {\bibfnamefont {J.-R.}\ \bibnamefont
  {Zhang}},\ }\href {\doibase 10.1103/PhysRevD.103.074016} {\bibfield
  {journal} {\bibinfo  {journal} {Phys. Rev. D}\ }\textbf {\bibinfo {volume}
  {103}},\ \bibinfo {pages} {074016} (\bibinfo {year} {2021})},\ \Eprint
  {http://arxiv.org/abs/2011.04594} {arXiv:2011.04594 [hep-ph]} \BibitemShut
  {NoStop}%
\bibitem [{\citenamefont {Kawanai}\ and\ \citenamefont
  {Sasaki}(2012)}]{Kawanai:2011jt}%
  \BibitemOpen
  \bibfield  {author} {\bibinfo {author} {\bibfnamefont {T.}~\bibnamefont
  {Kawanai}}\ and\ \bibinfo {author} {\bibfnamefont {S.}~\bibnamefont
  {Sasaki}},\ }\href {\doibase 10.1103/PhysRevD.85.091503} {\bibfield
  {journal} {\bibinfo  {journal} {Phys. Rev. D}\ }\textbf {\bibinfo {volume}
  {85}},\ \bibinfo {pages} {091503} (\bibinfo {year} {2012})},\ \Eprint
  {http://arxiv.org/abs/1110.0888} {arXiv:1110.0888 [hep-lat]} \BibitemShut
  {NoStop}%
\bibitem [{\citenamefont {Yang}\ \emph
  {et~al.}(2021{\natexlab{b}})\citenamefont {Yang}, \citenamefont {Ping},\ and\
  \citenamefont {Segovia}}]{Yang:2021hrb}%
  \BibitemOpen
  \bibfield  {author} {\bibinfo {author} {\bibfnamefont {G.}~\bibnamefont
  {Yang}}, \bibinfo {author} {\bibfnamefont {J.}~\bibnamefont {Ping}}, \ and\
  \bibinfo {author} {\bibfnamefont {J.}~\bibnamefont {Segovia}},\ }\href
  {\doibase 10.1103/PhysRevD.104.014006} {\bibfield  {journal} {\bibinfo
  {journal} {Phys. Rev. D}\ }\textbf {\bibinfo {volume} {104}},\ \bibinfo
  {pages} {014006} (\bibinfo {year} {2021}{\natexlab{b}})},\ \Eprint
  {http://arxiv.org/abs/2104.08814} {arXiv:2104.08814 [hep-ph]} \BibitemShut
  {NoStop}%
\bibitem [{\citenamefont {Hiyama}\ \emph {et~al.}(2003)\citenamefont {Hiyama},
  \citenamefont {Kino},\ and\ \citenamefont {Kamimura}}]{Hiyama:2003cu}%
  \BibitemOpen
  \bibfield  {author} {\bibinfo {author} {\bibfnamefont {E.}~\bibnamefont
  {Hiyama}}, \bibinfo {author} {\bibfnamefont {Y.}~\bibnamefont {Kino}}, \ and\
  \bibinfo {author} {\bibfnamefont {M.}~\bibnamefont {Kamimura}},\ }\href
  {\doibase 10.1016/S0146-6410(03)90015-9} {\bibfield  {journal} {\bibinfo
  {journal} {Prog. Part. Nucl. Phys.}\ }\textbf {\bibinfo {volume} {51}},\
  \bibinfo {pages} {223} (\bibinfo {year} {2003})}\BibitemShut {NoStop}%
\bibitem [{\citenamefont {Aguilar}\ and\ \citenamefont
  {Combes}(1971)}]{Aguilar:1971ve}%
  \BibitemOpen
  \bibfield  {author} {\bibinfo {author} {\bibfnamefont {J.}~\bibnamefont
  {Aguilar}}\ and\ \bibinfo {author} {\bibfnamefont {J.~M.}\ \bibnamefont
  {Combes}},\ }\href {\doibase 10.1007/BF01877510} {\bibfield  {journal}
  {\bibinfo  {journal} {Commun. Math. Phys.}\ }\textbf {\bibinfo {volume}
  {22}},\ \bibinfo {pages} {269} (\bibinfo {year} {1971})}\BibitemShut
  {NoStop}%
\bibitem [{\citenamefont {Balslev}\ and\ \citenamefont
  {Combes}(1971)}]{Balslev:1971vb}%
  \BibitemOpen
  \bibfield  {author} {\bibinfo {author} {\bibfnamefont {E.}~\bibnamefont
  {Balslev}}\ and\ \bibinfo {author} {\bibfnamefont {J.~M.}\ \bibnamefont
  {Combes}},\ }\href {\doibase 10.1007/BF01877511} {\bibfield  {journal}
  {\bibinfo  {journal} {Commun. Math. Phys.}\ }\textbf {\bibinfo {volume}
  {22}},\ \bibinfo {pages} {280} (\bibinfo {year} {1971})}\BibitemShut
  {NoStop}%
\bibitem [{\citenamefont {B.}(1972)}]{Simon1972}%
  \BibitemOpen
  \bibfield  {author} {\bibinfo {author} {\bibfnamefont {S.}~\bibnamefont
  {B.}},\ }\href {\doibase 10.1007/BF01649654} {\bibfield  {journal} {\bibinfo
  {journal} {Commun. Math. Phys.}\ }\textbf {\bibinfo {volume} {27}},\ \bibinfo
  {pages} {1} (\bibinfo {year} {1972})}\BibitemShut {NoStop}%
\bibitem [{\citenamefont {Ho}(1983)}]{HO19831}%
  \BibitemOpen
  \bibfield  {author} {\bibinfo {author} {\bibfnamefont {Y.}~\bibnamefont
  {Ho}},\ }\href {\doibase https://doi.org/10.1016/0370-1573(83)90112-6}
  {\bibfield  {journal} {\bibinfo  {journal} {Physics Reports}\ }\textbf
  {\bibinfo {volume} {99}},\ \bibinfo {pages} {1} (\bibinfo {year}
  {1983})}\BibitemShut {NoStop}%
\bibitem [{\citenamefont {Eichten}\ \emph {et~al.}(1975)\citenamefont
  {Eichten}, \citenamefont {Gottfried}, \citenamefont {Kinoshita},
  \citenamefont {Kogut}, \citenamefont {Lane},\ and\ \citenamefont
  {Yan}}]{Eichten:1974af}%
  \BibitemOpen
  \bibfield  {author} {\bibinfo {author} {\bibfnamefont {E.}~\bibnamefont
  {Eichten}}, \bibinfo {author} {\bibfnamefont {K.}~\bibnamefont {Gottfried}},
  \bibinfo {author} {\bibfnamefont {T.}~\bibnamefont {Kinoshita}}, \bibinfo
  {author} {\bibfnamefont {J.~B.}\ \bibnamefont {Kogut}}, \bibinfo {author}
  {\bibfnamefont {K.~D.}\ \bibnamefont {Lane}}, \ and\ \bibinfo {author}
  {\bibfnamefont {T.-M.}\ \bibnamefont {Yan}},\ }\href {\doibase
  10.1103/PhysRevLett.34.369} {\bibfield  {journal} {\bibinfo  {journal} {Phys.
  Rev. Lett.}\ }\textbf {\bibinfo {volume} {34}},\ \bibinfo {pages} {369}
  (\bibinfo {year} {1975})},\ \bibinfo {note} {[Erratum: Phys.Rev.Lett. 36,
  1276 (1976)]}\BibitemShut {NoStop}%
\bibitem [{\citenamefont {Eichten}\ \emph {et~al.}(1980)\citenamefont
  {Eichten}, \citenamefont {Gottfried}, \citenamefont {Kinoshita},
  \citenamefont {Lane},\ and\ \citenamefont {Yan}}]{Eichten:1979ms}%
  \BibitemOpen
  \bibfield  {author} {\bibinfo {author} {\bibfnamefont {E.}~\bibnamefont
  {Eichten}}, \bibinfo {author} {\bibfnamefont {K.}~\bibnamefont {Gottfried}},
  \bibinfo {author} {\bibfnamefont {T.}~\bibnamefont {Kinoshita}}, \bibinfo
  {author} {\bibfnamefont {K.~D.}\ \bibnamefont {Lane}}, \ and\ \bibinfo
  {author} {\bibfnamefont {T.-M.}\ \bibnamefont {Yan}},\ }\href {\doibase
  10.1103/PhysRevD.21.203} {\bibfield  {journal} {\bibinfo  {journal} {Phys.
  Rev. D}\ }\textbf {\bibinfo {volume} {21}},\ \bibinfo {pages} {203} (\bibinfo
  {year} {1980})}\BibitemShut {NoStop}%
\bibitem [{\citenamefont {Zhao}\ \emph {et~al.}(2020)\citenamefont {Zhao},
  \citenamefont {Zhou}, \citenamefont {Chen},\ and\ \citenamefont
  {Zhuang}}]{Zhao:2020jqu}%
  \BibitemOpen
  \bibfield  {author} {\bibinfo {author} {\bibfnamefont {J.}~\bibnamefont
  {Zhao}}, \bibinfo {author} {\bibfnamefont {K.}~\bibnamefont {Zhou}}, \bibinfo
  {author} {\bibfnamefont {S.}~\bibnamefont {Chen}}, \ and\ \bibinfo {author}
  {\bibfnamefont {P.}~\bibnamefont {Zhuang}},\ }\href {\doibase
  10.1016/j.ppnp.2020.103801} {\bibfield  {journal} {\bibinfo  {journal} {Prog.
  Part. Nucl. Phys.}\ }\textbf {\bibinfo {volume} {114}},\ \bibinfo {pages}
  {103801} (\bibinfo {year} {2020})},\ \Eprint
  {http://arxiv.org/abs/2005.08277} {arXiv:2005.08277 [nucl-th]} \BibitemShut
  {NoStop}%
\bibitem [{\citenamefont {Yang}\ \emph
  {et~al.}(2020{\natexlab{c}})\citenamefont {Yang}, \citenamefont {Ping},
  \citenamefont {Ortega},\ and\ \citenamefont {Segovia}}]{Yang:2019lsg}%
  \BibitemOpen
  \bibfield  {author} {\bibinfo {author} {\bibfnamefont {G.}~\bibnamefont
  {Yang}}, \bibinfo {author} {\bibfnamefont {J.}~\bibnamefont {Ping}}, \bibinfo
  {author} {\bibfnamefont {P.~G.}\ \bibnamefont {Ortega}}, \ and\ \bibinfo
  {author} {\bibfnamefont {J.}~\bibnamefont {Segovia}},\ }\href {\doibase
  10.1088/1674-1137/44/2/023102} {\bibfield  {journal} {\bibinfo  {journal}
  {Chin. Phys. C}\ }\textbf {\bibinfo {volume} {44}},\ \bibinfo {pages}
  {023102} (\bibinfo {year} {2020}{\natexlab{c}})},\ \Eprint
  {http://arxiv.org/abs/1904.10166} {arXiv:1904.10166 [hep-ph]} \BibitemShut
  {NoStop}%
\bibitem [{\citenamefont {Yang}\ and\ \citenamefont
  {Ping}(2018)}]{Yang:2017rpg}%
  \BibitemOpen
  \bibfield  {author} {\bibinfo {author} {\bibfnamefont {G.}~\bibnamefont
  {Yang}}\ and\ \bibinfo {author} {\bibfnamefont {J.}~\bibnamefont {Ping}},\
  }\href {\doibase 10.1103/PhysRevD.97.034023} {\bibfield  {journal} {\bibinfo
  {journal} {Phys. Rev.}\ }\textbf {\bibinfo {volume} {D97}},\ \bibinfo {pages}
  {034023} (\bibinfo {year} {2018})},\ \Eprint
  {http://arxiv.org/abs/1703.08845} {arXiv:1703.08845 [hep-ph]} \BibitemShut
  {NoStop}%
\bibitem [{\citenamefont {Yang}\ \emph
  {et~al.}(2020{\natexlab{d}})\citenamefont {Yang}, \citenamefont {Ping},\ and\
  \citenamefont {Segovia}}]{gy:2020dhts}%
  \BibitemOpen
  \bibfield  {author} {\bibinfo {author} {\bibfnamefont {G.}~\bibnamefont
  {Yang}}, \bibinfo {author} {\bibfnamefont {J.}~\bibnamefont {Ping}}, \ and\
  \bibinfo {author} {\bibfnamefont {J.}~\bibnamefont {Segovia}},\ }\href
  {\doibase 10.1103/PhysRevD.102.054023} {\bibfield  {journal} {\bibinfo
  {journal} {Phys. Rev. D}\ }\textbf {\bibinfo {volume} {102}},\ \bibinfo
  {pages} {054023} (\bibinfo {year} {2020}{\natexlab{d}})}\BibitemShut
  {NoStop}%
\bibitem [{\citenamefont {Yang}\ \emph
  {et~al.}(2020{\natexlab{e}})\citenamefont {Yang}, \citenamefont {Ping},\ and\
  \citenamefont {Segovia}}]{gy:2020dht}%
  \BibitemOpen
  \bibfield  {author} {\bibinfo {author} {\bibfnamefont {G.}~\bibnamefont
  {Yang}}, \bibinfo {author} {\bibfnamefont {J.~L.}\ \bibnamefont {Ping}}, \
  and\ \bibinfo {author} {\bibfnamefont {J.}~\bibnamefont {Segovia}},\ }\href
  {\doibase 10.1103/PhysRevD.101.014001} {\bibfield  {journal} {\bibinfo
  {journal} {Phys. Rev. D}\ }\textbf {\bibinfo {volume} {101}},\ \bibinfo
  {pages} {014001} (\bibinfo {year} {2020}{\natexlab{e}})}\BibitemShut
  {NoStop}%
\end{thebibliography}%

\end{document}